\documentclass[aps,pra,twocolumn,,superscriptaddress,floatfix]{revtex4-1}
\usepackage[svgnames]{xcolor}
\definecolor{darkred}{rgb}{0.6, 0, 0}
\definecolor{darkgreen}{rgb}{0, 0.5, 0}
\usepackage[colorlinks=true, urlcolor = darkgreen, linkcolor = darkgreen, citecolor = teal, bookmarks=false]{hyperref}
\usepackage{amsfonts,amsmath,amssymb,amsthm}
\usepackage{graphicx}
\usepackage{dsfont}
\usepackage[normalem]{ulem}
\usepackage{enumerate}
\usepackage{natbib}

\usepackage{tcolorbox}
\usepackage{tabularx}
\usepackage{array}
\usepackage{colortbl}
\tcbuselibrary{skins}

\swapnumbers

\def\ii{{\rm i}}
\newcommand{\dd}{{\rm d}}

\def\expect#1{\langle#1\rangle}

\def\ol#1{\bar{#1}}

 \newcommand{\Jr}{\mathrm{J}}
\newcommand{\jr}{\mathrm{j}}
\newcommand{\Jcal}{\mathcal{J}}
\newcommand{\qr}{\mathrm{q}}

\begin{document}

\title{
Exact anomalous current fluctuations in a deterministic interacting model
}

\author{\v{Z}iga Krajnik}
\affiliation{Faculty for Mathematics and Physics,
University of Ljubljana, Jadranska ulica 19, 1000 Ljubljana, Slovenia}

\author{Johannes Schmidt}
\affiliation{Technische Universit\"at Berlin, Institute for Theoretical Physics, Hardenbergstr. 36, D-10623 Berlin, Germany}
\affiliation{Bonacci GmbH, Robert-Koch-Str. 8, 50937 Cologne, Germany}

\author{Vincent Pasquier}
\affiliation{Institut de Physique Th\'{e}orique, Universit\'{e} Paris Saclay, CEA, CNRS UMR 3681, 91191
Gif-sur-Yvette, France}

\author{Enej Ilievski}
\affiliation{Faculty for Mathematics and Physics,
University of Ljubljana, Jadranska ulica 19, 1000 Ljubljana, Slovenia}

\author{Toma\v{z} Prosen}
\affiliation{Faculty for Mathematics and Physics,
University of Ljubljana, Jadranska ulica 19, 1000 Ljubljana, Slovenia}

\date{\today}

\begin{abstract}
We analytically compute the full counting statistics of charge transfer in a classical automaton of interacting charged particles.
Deriving a closed-form expression for the moment generating function with respect to a stationary equilibrium state,
we employ asymptotic analysis to infer the structure of charge current fluctuations for a continuous range of timescales.
The solution exhibits several unorthodox features. Most prominently, on the timescale of typical fluctuations
the probability distribution of the integrated charge current in a stationary ensemble without bias is distinctly non-Gaussian
despite diffusive behavior of dynamical charge susceptibility. While inducing a charge imbalance is enough to recover Gaussian
fluctuations, we find that higher cumulants grow indefinitely in time with different exponents, implying singular scaled cumulants. We associate this phenomenon with the lack of a regularity condition on moment generating functions and the
onset of a dynamical critical point. In effect, the scaled cumulant generating function does not, irrespectively of charge bias, represent a faithful generating function of the scaled cumulants, yet the associated Legendre dual yields the correct large-deviation rate function.
Our findings hint at novel types of dynamical universality classes in deterministic many-body systems.
\end{abstract}

\pacs{02.30.Ik,05.70.Ln,75.10.Jm}

\maketitle

{\bf Introduction.}---The central limit theorem (CLT) is one of the bedrock accomplishments of probability theory.
In the standard formulation, the CLT asserts that sums of random, independent, identically distributed variables converge towards the normal distribution when the sample size becomes large. Validity of the CLT however transcends uncorrelated processes, as it applies
for macroscopic fluctuating observables in a wide array of dynamical processes in nature, including classical or quantum deterministic dynamical systems which typically exhibit highly non-trivial temporal correlations.
It appears as though the CLT only ceases to hold away from equilibrium, i.e. upon breaking reversibility at the microscopic level. 

Another hallmark result of statistical analysis is the large deviation principle (LDP) \cite{Ellis_book,DemboZeitouni_book,Touchette_LDT},
stipulating that atypically large (rare) fluctuations are exponentially unlikely.
In this regard, the main object of interest is a dynamical partition sum, the moment generating function (MGF) of the process
also known as the full counting statistics (FCS). The rate function describing large deviations can be inferred from the logarithm of MGF.
In spite of many important cases where MGF can be computed explicitly~\cite{Kipnis82,Derrida07,Derrida09,BD13,Znidaric14_LD,BD16,Yoshimura18,Moriya19,Gamayun20,MBHD20,DoyonMyers20}, threre are virtually
no explicit results available when it comes to genuinely \emph{interacting} many-particle systems governed by deterministic and reversible microscopic evolution laws, whether in or out of equilibrium.

A recent numerical study \cite{Krajnik_breakdown} has found robust signature of anomalous dynamical fluctuations
in the {\em integrable} Landau--Lifshitz ferromagnet, hinting that lack of ergodicity can play a pivotal role and may lead to
 inapplicability of the CLT. The precise microsopic mechanism leading to such an unconventional behavior has not been identified however.
In this letter, we report major progress on this question. We compute the exact FCS for a simple model of \emph{interacting} charged degrees
of freedom governed by a reversible deterministic equation of motion in a stationary equilibrium state.
By deducing the late-time behavior of cumulants in a closed analytic form, we encounter two novel regimes of dynamical behavior
characterized by divergent scaled cumulants of transferred charge.


{\bf Current fluctuations on typical and large scale.}---We consider an infinitely extended deterministic dynamical many-body system
with charge conservation. The time-integrated current density, $\Jr(t) = \int^{t}_{0}\dd \tau\,  \jr (\tau)$, where $\jr (\tau)$
is the charge-current density (at the origin) propagated by time $\tau$, can be viewed as a dynamical
fluctuating observable, measuring the net transferred charge between two halves of the system in the time interval $t$ for
each particular initial configuration. Obtaining the FCS of $\Jr (t)$ amounts to computing the MGF \cite{Touchette_LDT,Esposito_review}
\vspace{-0.1cm}
\begin{equation}
G(\lambda|t) \equiv \left\langle e^{\lambda\,\Jr (t)} \right\rangle
\equiv   \int \dd \Jr \,\mathcal{P}(\Jr|t)e^{\lambda \Jr},
\label{eqn:MGF}
\vspace{-0.1cm}
\end{equation}
corresponding to a Laplace transformation of the normalized (time-dependent) current distribution $\mathcal{P}(\Jr|t)$
of $\Jr(t)$, computed with respect to a stationary equilibrium measure. The formal variable $\lambda \in {\mathbb{C}}$ is commonly
known as fugacity (or counting field). Owing to detailed balance we have the symmetry $\mathcal{P}(\Jr |t)=\mathcal{P}(-\Jr|t)$,
implying $G(\lambda|t)=G(-\lambda|t)$.
The associated cumulant generating function (CGF), $\log G(\lambda|t) = \sum_{n=0}^\infty c_n(t) \frac{\lambda^n}{n!}$,
encodes an entire hierarchy of connected $n$-point dynamical correlation functions of time-integrated current densities
\begin{equation}
c_{n}(t) = \int^{t}_{0}  \prod_{k=1}^{n} d\tau_{k}\expect{\jr(\tau_{n})\jr(\tau_{n-1})\cdots \jr(\tau_{1})}^{c}.
\label{eqn:cumulants}
\end{equation}
Assuming that, for asymptotically large times, the MGF grows as $G(\lambda|t)\asymp \exp{(t^{\alpha}F(\lambda))}$,
we introduce the \emph{scaled} CGF
\begin{equation}
F(\lambda|t)\equiv t^{-\alpha} \log G(\lambda|t),\quad F(\lambda) \equiv \lim_{t\to \infty} F(\lambda|t),
\label{eqn:dynamical_free_energy}
\end{equation}
which may be viewed as a `dynamical free energy' of the process. We stress that exponent $\alpha$ is intrinsic to the system.
In nonequilibrium processes, such as current-carrying steady-states arising in the boundary-driven systems in 
stochastic systems with an intrinsic asymmetric drift, one finds ballistic scaling with $\alpha=1$. For ergodic diffusive systems
in equilibrium, such as the simple symmetric exclusion processes \cite{Derrida09}, one instead has $\alpha=1/2$.


In this study, we focus exclusively on \emph{equilibrium} states where, by detailed balance, odd moments vanish,
$\expect{[\Jr (t)]^{2n-1}}=0$, while variance $\expect{[\Jr(t)]^{2}}^{c} \sim t^{1/z}$
sets the scale of \emph{typical} fluctuations $\Jr(t)\sim t^{1/2z}$ governed by the dynamical exponent $z\geq 1$.
Probabilities of atypically large fluctuations on a timescale {$t^\zeta$}, in the range $1/2z<\zeta\leq 1$,
can be quantified in terms of the rescaled current $\Jcal(t)=t^{-\zeta}\Jr(t)$.
Assuming that the probabilities of measuring atypical values of the current $\Jcal$ are
exponentially suppressed, we anticipate that
\begin{equation}
\mathbb{P} \big(\Jcal(t)= j\big) \asymp \exp{\left[-t^{v(\zeta)}I_{\zeta}(j)\right]},
\label{eqn:LD_form}
\end{equation}
where $I_{\zeta}(j)$, with ${j}\equiv \lim_{t\to \infty}t^{-\zeta}\Jr(t)$,
 is the LD \emph{rate function},
and $v(\zeta)$ the associated `speed' that depends, in general, on the adjustable scale parameter $\zeta$.
In the context of LD theory, one is typically interested in the \emph{largest} deviations corresponding to a
ballistic scaling exponent $\zeta=1$, with the associated speed $v(\zeta)=\zeta$.
Fluctuations within the range of scales $1/2z<\zeta<1$ are commonly referred to as `moderate deviations'.
Scaled CGF $F(\lambda)$ is not just a formal object: for $\lambda\in\mathbb R$ it represents a convex function that
takes a distinguished role in LD theory. Provided $F(\lambda)$ is everywhere differentiable on its domain,
the G\"{a}rtner--Ellis theorem \cite{Touchette_LDT} states that its Legendre transform,
$I(j)={\rm max}_{\lambda}[\lambda\, {j}-F(\lambda)]$ provides the rate function
$I( j)\equiv I_{\zeta=1}({j})$, quantifying probabilities of exponentially rare events.

Now we touch a delicate but pivotal point. In the literature devoted to applications of LD theory,
it is commonly understood that $F(\lambda)$ provides the generating series for the \emph{scaled} cumulants
$s_{n} = \lim_{t\to \infty}t^{-\alpha}c_{n}(t)$ when expanded around $\lambda=0$,
$s_{n}=(\dd/\dd \lambda)^{n}F(\lambda)\big|_{\lambda=0}$. At the technical level, however,
faithfulness of $F(\lambda)$ hinges on interchangeabilty of the limit $t\to \infty$ and
the series expansion of $F(\lambda|t)$. Indeed, there is no reason to apriori assume
(i) that the sequence of analytic function $F(\lambda|t)$ converges necessarily to an analytic limiting function $F(\lambda)$ 
 nor (ii) that, assuming $F(\lambda)$ is analytic, its expansion coefficients yield
scaled cumulants $s_{n}$. As we argue next, not only are both of these scenarios viable, but they lead to
certain profound physical consequences.

{\bf Evading the Central Limit Theorem.}---Before considering our model, we explain how robustness of the CLT is in fact
deeply rooted in analytic properties of $G(\lambda|t)$. Whenever there exists a disc ${\rm D}(r)$ of radius $r>0$ centered at $\lambda=0$,
such that $F(\lambda|t)$ are uniformly bounded on ${\rm D}(r)$ for all times $t$, and $F(\lambda)$ exists, then --- proven in a theorem by Bryc \cite{Bryc93} assuming $\alpha=1$ (see also \cite{Jaksic12}) --- the central limit theorem holds as a consequence of \emph{finite} scaled cumulants $s_{n}$ (ensured by Vitali's convergence theorem).
To put it simply, an irregular (i.e. non-generic) behaviour can only be achieved when Bryc's analyticity conditions are not fulfilled. 

With this in mind, we now imagine a system of interacting ballistically propagating quasiparticles, where it is expected
(e.g. based on generalized hydrodynamics \cite{MBHD20,DoyonPerfeto}) that MGF, cf. Eq.~\eqref{eqn:MGF}, grows asymptotically with dyamical exponent $\alpha=1$. If the model admits a $\mathbb{Z}_{2}$ parity symmetry (e.g. charge conjugation) under which the charge current flips sign then,
by analogy with quantum spin chains \cite{PhysRevLett.111.057203,IN17,transport_review},
the charge Drude weight (in a parity-invariant equilibrium state) identically vanishes,
and charge transport is governed by a \emph{subballsitic} dynamical exponent $z>1$.
Recalling that $c_{2}(t)\sim t^{1/z}$, in this scenario scaled cumulants $s_{n}$ \emph{cannot} correspond to coefficients
of $F(\lambda)$ (in fact, $s_{n}$ need not even exist). Interestingly however, as we are about to demonstrate next, even
ballsitic charge transport (i.e. $\alpha=z=1$) does not by itself guarantee faithfulness of $F(\lambda)$.
To substantiate our claims, we explicitly compute the FCS for the charge-current fluctuations in a 
classical deterministic cellular automaton of hard-core interacting  charged particles introduced and studied earlier
in \cite{Medenjak17,Klobas2018,Medenjak19}. We first give a closed-from solution for the FCS and subsequently discuss its most salient features.

\begin{figure}[htb]
\centering
\includegraphics[width=0.45\textwidth]{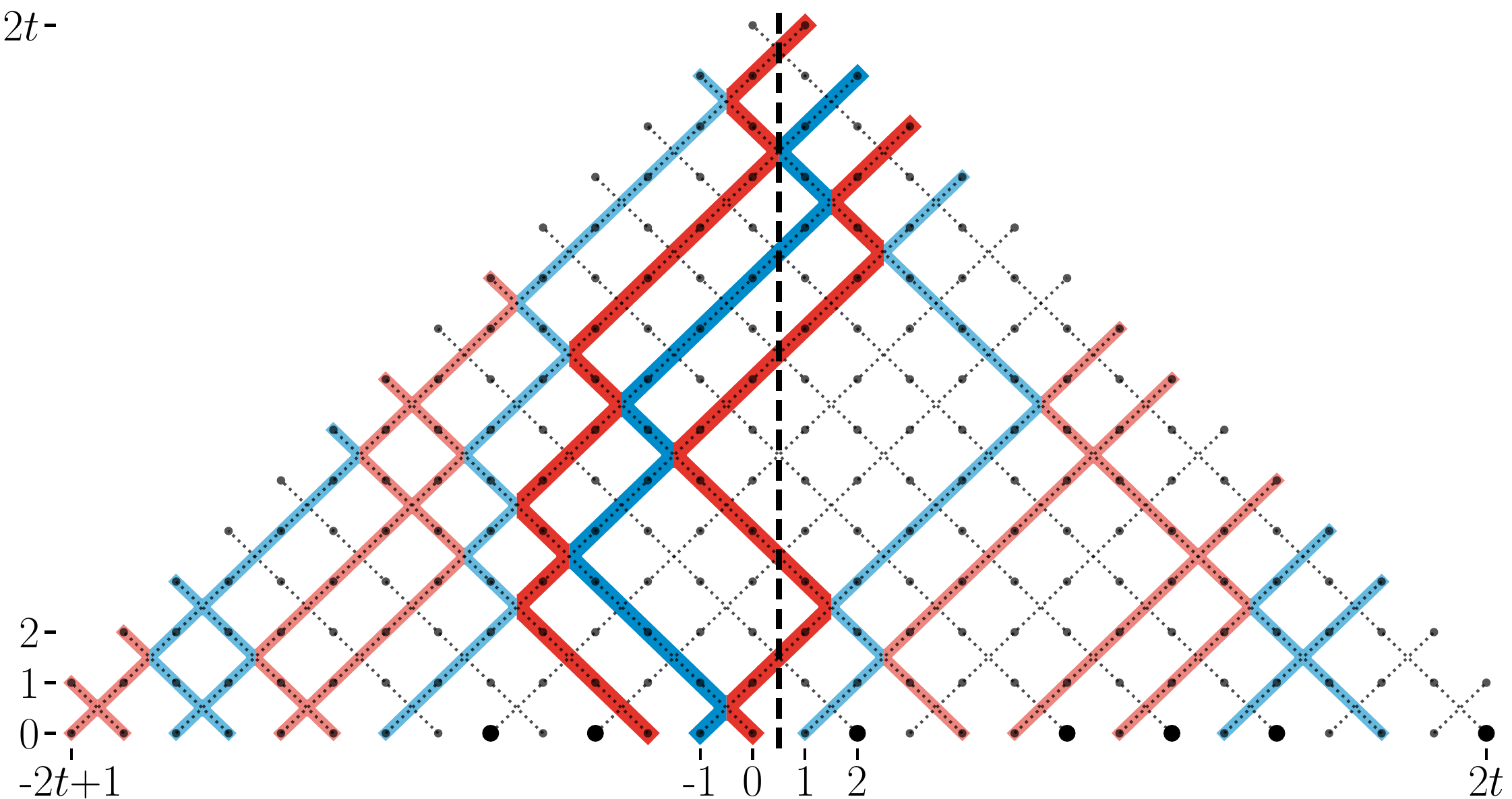}
\caption{Coordinate frame (time vertical, space horizontal) of a deterministic
charged hard-core lattice gas (red: $+$ particles, blue: $-$ particles, while thin black
lines indicate vacancies). Example of a lightcone (pyramid) section of a typical trajectory, for
which initial data on a saw of $4t$ subsequent links uniquely determine the transport
through the mid-point (dashed line) for all times times from $0$ to $2t$.
}
\label{fig1}
\vspace{-0.5cm}
\end{figure}

{\bf Cellular automaton with solvable FCS.}---We consider a reversible cellular automaton introduced in \cite{Medenjak17},
realized as a space-time circuit composed of elementary two-body maps. Each lattice site $(\ell,t)\in \mathbb{Z}\times \mathbb{Z}$
is occupied with one of three `species' of particles taking values in $\mathcal{Q}=\{\emptyset,+,-\}$, representing a particle of positive ($+$) or negative charge ($-$), and a charge-neutral vacancy ($\emptyset$).
The local two-particle propagator $\Phi:\mathcal{Q}\times \mathcal{Q}\to \mathcal{Q}\times \mathcal{Q}$
acts simply as a permutation (i.e. no interaction) whenever the sum of charges is non-zero,
$(\emptyset,\qr)\leftrightarrow(\qr,\emptyset)$ for $\qr \in \{\emptyset,+,-\}$,
while oppositely charged particles repel each other and thus retain their initial positions,
$(\qr,\qr')\leftrightarrow (\qr,\qr')$ for $\qr,\qr'\in \{\pm\}$. The dynamics consists of brickwork application of $\Phi$,
as depicted in Figure~\ref{fig1}.

Dynamical properties of such an automaton are distinctly non-ergodic; a phase-space trajectory cannot explore the entire phase space: besides conservation of total charge, it possesses an exponentially large number of conserved quantities (related trivially to the
fact that vacancies are inert). The exact charge Drude weight and the diffusion constant have been computed in \cite{Klobas2018}.

Our central result is an explicit computation of the MGF which is outlined below.
The ensemble average in $G(\lambda|t)$ can be computed in a two-stage `nested' way: for every `frozen' sublattice $\Sigma\subset \mathbb Z$
of occupied sites at $t=0$ we perform an
average over all possible charge configurations $\{q_\ell\}$, and subsequently average over all sublattices $\Sigma$. Let $\Lambda_\pm \subset \Sigma$ denote the sublattices of charged particles at $t=0$ that move from the right to the left half (respectively, vice-versa) during time interval $[0, 2t]$.
The integrated current through the origin corresponds to the total transferred charge,
$\Jr(t)=\sum_{\ell \in \Lambda_{-}}\qr_{\ell}-\sum_{\ell \in \Lambda_{+}}\qr_{\ell}$.
We make the following observations: (a) particle world-lines cannot cross each other, implying in particular that at most one of the subsets $\Lambda_{\pm}$ can be non-empty, (b) the signed number of worldlines crossing the origin is given by the difference between the number of vacancies passing through the origin from the left/right up to time $2t$.
Introducing a separable invariant probability measure $\mathbb{P}(\{\qr_{\ell}\})=\prod_{\ell}p(\qr_{\ell})$, with
$p(\pm)=\tfrac{1}{2}\rho(1\pm b)$, $p(\emptyset)=\ol{\rho}=1-\rho$ corresponding to densities of particles and vacancies,
we derived (see \cite{suppmat} for details) the following exact double-sum representation for the MGF
\begin{equation}
G(\lambda|t) = \! \sum_{l,r=0}^{t}\binom{t}{l}\binom{t}{r}\ol{\rho}^{l+r}\rho^{2t-l-r}
\prod_{\varepsilon=\pm}[\mu_{\varepsilon}(\lambda)]^{d_{\varepsilon}(l,r)},
\label{eqn:exact_MGF}
\end{equation}
where $\mu_{\pm}(\lambda)\equiv \cosh{(\lambda)} \mp b\sinh{(\lambda)}$, $d_{\varepsilon}(l,r)\equiv (|l-r|+\varepsilon(l-r))/2$,
and $b\in [0,1)$ is the `charge bias'. 
In the following, we systematically carry out an asymptotic analysis of $G(\lambda|t)$.

{\bf Dynamical free energy and cumulants.}---Performing asymptotic analysis on $F(\lambda|t)$, see Eq.~\eqref{eqn:dynamical_free_energy},
for $\alpha=1$, $\lambda\in\mathbb R$, we inferred the following scaled CGF
\begin{equation}
F(\lambda) = \log\left[1+\varDelta^{2}\Big(\mu_{b}+\mu^{-1}_{b}-2\Big)\right],
\label{eqn:F}
\end{equation}
with $\mu_{b}\equiv \cosh \lambda+|b|\sinh{|\lambda|}$
and $\varDelta^2 \equiv \rho(1-\rho) \in [0,\tfrac{1}{4}]$.
We stress, importantly, that $F(\lambda)$ does not provide (irrespectively of $b$) the generating series for scaled cumulants $s_{n}$.
As already announced, we find (regardless of $b$) that all scaled cumulants $s_{n>2}$ are \emph{singular}.
We deduced the following asymptotic behavior~\footnote{We use a notational convention where $A^{[b]}$ refers to a quantity $A$ at a generic value of the bias parameter $0<|b|<1$, while the cases for $b=0$, and $b=1$, are denoted, respectively as $A^{[0]}$, and $A^{[1]}$.}
\begin{equation}
c^{[b]}_{2n>2}(t) \sim t^{n-1/2},\quad \quad c^{[0]}_{2n}(t) \sim t^{n/2},
\label{eqn:cumulant_growth}
\end{equation}
and succeeded in obtaining a compact generator of cumulant asymptotics,
$c_n^{[b \geq 0]}(t) \asymp (\dd/\dd\lambda)^n\mathcal{F}^{[b \geq 0]}(\lambda)|_{\lambda=0}$ with $\exp \left[\mathcal{F}^{[b\geq0]}(\lambda)\right] = \sum_{\varepsilon = \pm} \exp({a_\varepsilon^2})\big(1+ \textrm{erf}(a_\varepsilon)\big)$, where
$a_{\pm} \equiv t^{1/2} \varDelta \left[\tfrac{1}{2}(1-b^2) \lambda^2 \pm b \lambda \right]$ for $b\in [0,1)$.

{\bf Typical fluctuations and CLT.}---We first examine fluctuations of $\Jr (t)$ on the `typical timescale', associated with
scaling exponent $\zeta =1/2z$. To this end, we explicitly compute cumulants $\kappa_{n}(t)$ characterizing the
time-dependent distribution $\mathcal{P}_{1/2z}(\Jcal|t)$.
Recalling a theorem by Marcinkiewicz \cite{Marcinkiewicz}, stating that the Gaussian distribution is the unique distribution with finitely many non-zero cumulants, the CLT 
applies if and only if $\lim_{t\to \infty}\kappa_{n}(t)=0$ for all $n>2$.
From the scaling relation $\kappa_{n}(t) = t^{-n/2z}c_{n}(t)$ we readily deduce the scalings
$\kappa^{[b]}_{n>2}(t) \sim t^{-1/2}$ and $\kappa^{[0]}_{n>2}(t) \sim t^{0}$.
For a finite bias $b>0$, all the higher cumulants $\kappa_{n>2}^{[b]}(t)$ of the distribution $\mathcal{P}^{[b]}_{1/2z}(\Jcal|t)$ decay with time, yielding a Gaussian asymptotic profile, ${\mathcal{P}_{\rm typ}(j) \equiv \lim_{t \to \infty} \mathcal{P}_{\zeta_{\rm typ}}(\Jcal=j|t)}$,
\begin{equation}
\mathcal{P}^{[b]}_{\rm typ}(j)=\frac{1}{\sqrt{2\pi \sigma^{2}}}\exp{\left[-\frac{j^{2}}{2\sigma^{2}}\right]},
\quad \sigma^{2}=2(b\varDelta)^{2}.
\label{eqn:typical_Gaussian}
\end{equation}
Upon switching off the bias, $b=0$, typical fluctuations occur on a scale $\zeta=1/4$, where we inferred
a \emph{non-Gaussian} profile characterized by finite cumulants $\kappa^{[0]}_{n}=\lim_{t\to \infty}\kappa^{[0]}_{n}(t)<\infty$,
with the following integral representation \cite{suppmat}
\begin{equation}
\mathcal{P}^{[0]}_{\rm typ}({j}) = \frac{1}{\sqrt{2}\pi\varDelta}\int_{\mathbb{R}}\dd u
\exp{\left[-\Big(\frac{u^{2}}{2\varDelta}\Big)^{2}-\frac{ j^{2}}{2u^{2}}\right]}.
\label{eqn:typical_nonGaussian}
\end{equation}
Explicitly, $\mathcal{P}^{[0]}_{\rm typ}({j})=(2\varDelta)^{-1/2}M_{1/4}\big(\sqrt{2/\varDelta}\,|{j}|\big)$,
where $M_{\nu}(x)\equiv \sum_{k=0}^{\infty}(-x)^{k}/[k!\,\Gamma((1-\nu)-\nu\,k)]$
is the M-Wright function \cite{Mainardi_2010}.
The associated MGF reads explicitly
$\mathcal{G}^{[0]}(\eta)=e^{w^4}(1 + {\rm erf}\, w^{2}) = E_{1/2}(w^2)$,
where $w^2 = \eta^2 \varDelta/2$ and $E_{\nu}(x)=\sum_{k= 0}^{\infty}x^{k}/\Gamma(1+\nu\,k)$ is the Mittag-Leffler function
(note that $\mathcal{G}^{[0]}(\eta)=\exp{[\mathcal{F}^{[0]}(\eta\,t^{-1/4})]}$).

{\bf Large deviation principle.}--- What remains is to characterize fluctuations on the largest scale $\zeta=1$.
Since $F(\lambda)$ is strictly convex and differentiable in its entire domain $\lambda \in \mathbb{R}$ for all values of $b$,
the G\"{a}rtner--Ellis theorem ensures that the Legendre transform is involutive \cite{Touchette_LDT}, and that
$I(j)$ corresponds to a unique strictly convex and differentiable LD rate function $I({j})$
(whose Legendre transform yields back $F(\lambda)$).

In the general case with finite bias, the leading-order behavior near $\lambda=0$
reads $F^{[b]}(\lambda)=(b\varDelta)\lambda^{2}+\mathcal{O}(|\lambda|^{3})$,
implying a quadratic rate function for perturbatively small  $j$,
$I^{[b]}( j) = \left(\frac{ j}{2b\varDelta}\right)^{2} + \mathcal{O}(j^{4})$.
At $b=0$, the behavior is markedly different; owing to the absence of the leading order terms in SCGF,
$F^{[0]}(\lambda)=(\varDelta/2)^{2}\lambda^{4}+\mathcal{O}(\lambda^{6})$, we find
$I^{[0]}({j}) = \frac{3}{4}({j}^{2}/\varDelta)^{2/3} + \mathcal{O}( j^2)$.
Unlike $I^{[b]}({j})$, $I^{[0]}({j})$ is not twice differentiable at ${j}=0$.
On the other hand, at large $|\lambda|$ we have $F(\lambda)\sim |\lambda|$, implying that ${j}$ is confined within
the compact interval $[-1,1]$, cf.~\cite{Znidaric14_LD,Buca19}. This is a direct
manifestation of causality: owing to the fact that charges propagate with unit velocity and interact locally,
the maximal transferred charge in a time interval $t$ is upper-bounded by $t$.  The near-horizon behavior can be found analytically \cite{suppmat}.

We have also computed a family of rate functions associated to the `moderate deviation principle' for a continuous range of
timescales ($1/2z<\zeta<1$) \cite{suppmat}.


\begin{figure}[t]
\centering
\includegraphics[width=0.48\textwidth]{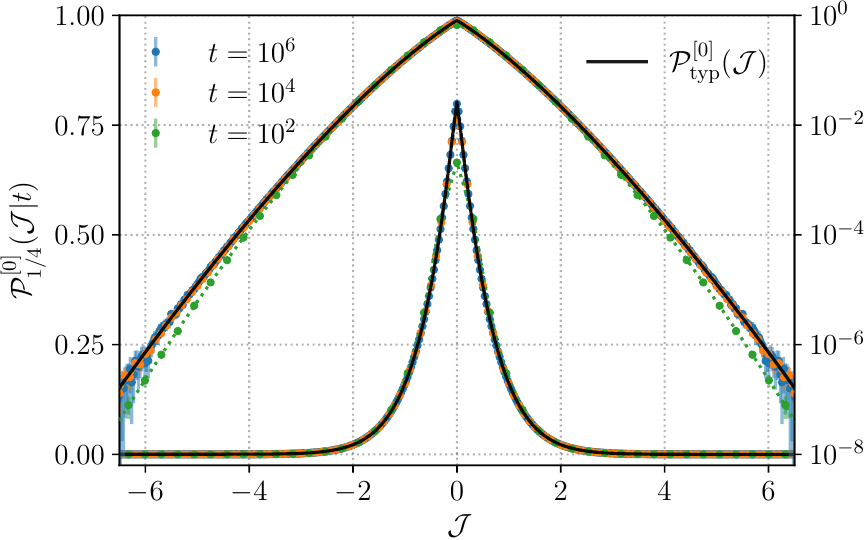}
\caption{Rescaled current distribution for unbiased $b=0$, half-filled $\rho=0.5$ charged hard-core lattice gas in normal/log scale (lower/upper data). Colored dashed lines show convergence of exact distributions $\mathcal{P}^{[0]}_{1/4}(\Jcal|t)$ to the asymptotic form 
(\ref{eqn:typical_nonGaussian}) (solid black line).
Estimated current distribution (dots), agrees with exact solution within statistical errors (${\rm N}_{\rm sample} = 10^9$).
}
\vspace{-0.5cm}
\end{figure}

{\bf Singular scaled cumulants and criticality.}---Lack of analyticity of scaled CGF $F(\lambda)$ is often found in Markovian stochastic systems driven away from equilibrium by means of boundary reservoirs, where it is attributed to a first-order dynamical phase transition (DPT), see Refs.~\cite{Garrahan07,FG13,Hickey13,Nyawo16,Nemoto17,Buca19,Jack20}. We are not aware of similar dynamical features
taking place in equilibrium. Despite that, we can observe certain conspicuous similarities.

Significance of divergent scaled cumulants is most transparently discussed in the complex fugacity plane
in the framework of the Lee--Yang theory \cite{LY52} of phase transitions \cite{BlytheEvans03,BENA_2005}. Presently, we find that
$\mathcal{O}(t)$ Lee--Yang zeros of $G(\lambda|t)$ condense along certain contours in the $\lambda$-plane.
By four-fold symmetry, there are four zeros of $G(\lambda|t)$ closest to the origin $\lambda=0$, at a distance $r(t)$, 
corresponding to the convergence radius of a complex Taylor series $\log G(\lambda|t)=\sum_{n}c_{n}(t) \lambda^{n}/n!$.
Applying the standard analysis (see Refs.~\cite{FG13,Brandner17,Deger18,Deger20}), and using the known asymptotics of $c_{n}(t)$,
we deduce the scaling $r^{[b]}(t)\sim t^{-1/2}$, $r^{[0]}(t) \sim t^{-1/4}$ (see \cite{suppmat} for details).
The vanishing convergence radius, $r_{\infty}\equiv\lim_{t\to \infty}r(t)=0$, signifies that $\lambda_{c}=0$ is a \emph{dynamical critical point}. 
Based on this, one might draw an incorrect conclusion that scaled CGF $F(\lambda)$ develops a non-analyticity at the critical point. 
In reality, only $F^{[b]}(\lambda)$ is found to be non-analytic, owing to the discontinuities
in its odd-order derivatives at the origin. Conversely, $F^{[0]}(\lambda)$, which depends on $\mu_{0}(\lambda)=\cosh{\lambda}$ and is derived via Eq.~\eqref{eqn:dynamical_free_energy}, represents a real analytic function; while its expansion coefficients are unrelated to cumulants, $F^{[0]}(\lambda)$ is the Legendre-dual of the LD rate function
$I^{[0]}({j})$.

In contrast to first-order DPTs seen in out-of-equilibrium stochastic processes
(where both the scaled CGF $F(\lambda)$ and LD rate function exhibit a cusp), we encounter, in the biased case $b>0$,
a cusp only in the second derivative, $(\dd/\dd \lambda)^{2} F^{[b]}(\lambda)$. This indicates, at a formal level,
a DPT of \emph{third order} at $\lambda = \lambda_c$, with the value at the cusp
being the dynamical charge-current susceptibility, $s_{2}=\lim_{t\to \infty}t^{-1}c_{2}(t)=\int^{t}_{0}\dd \tau\expect{\jr(\tau)\jr(0)}^{c}$. Note that result (\ref{eqn:exact_MGF}) can be reinterpreted as a Curie--Weiss like partition sum \cite{Flindt_PRR}, where $b$ plays the role of a magnetic field with a line of first order phase transitions at $b_c=0$, ending at $\lambda = \lambda_c$.

The Lee--Yang theory permits us to establish that divergent scaled cumulants, with an extra assumption that
$c_{n}(t)/c_{n+2}(t)\sim t^{-\gamma_n}$ with $\lim_{n\to\infty} \gamma_n > 0$,
imply $r_{\infty}=0$ (i.e. $\lambda_{c}=0$), and vice-versa. In this scenario `Bryc's regularity conditions' ensuring applicability of CLT are violated.
Indeed, in the present model Lebesgue's criterion of dominated convergence is not satisfied by the time-sequence of \emph{real} analytic functions $F(\lambda|t)$, irrespectively of bias $b$.
One should however be cautious, as neither divergent $s_{n}$ nor non-analytic $F(\lambda)$ automatically imply a departure from Gaussianity.
The fate of $\mathcal{P}_{\rm typ}({j})$ is instead predicated on the asymptotic scaling of the higher cumulants
$c_{n}(t)=(\dd/\dd \lambda)^{n}\log G(\lambda|t)|_{\lambda=0}$:
writing $c_{n>2}(t)\sim t^{\nu_{n}}$, one finds a Gaussian $\mathcal{P}_{\rm typ}({j})$
\emph{if and only if} the exponents $\nu_{n}$ can be upper-bounded by threshold exponents $\nu_{n}<n/2z$ and is otherwise \emph{violated}.

{\bf Fluctuations of particle current.}---It is instructive to add that fluctuations of the total transferred particle number behave regularly.
By disregarding internal charge degrees of freedom, the model reduces to free ballistically propagating particles with $z=\alpha=1$.
Setting accordingly $b=1$, the MGF can be easily summed up explicitly, yielding
an analytical (and faithful) scaled CGF of the form $F^{[1]}(\lambda)=\log[1+2\varDelta^{2}(\cosh{(\lambda)}-1)]$,
expectedly recovering the celebrated Levitov--Lesovik formula \cite{Levitov_book,Schonhammer07} (here specialized to a single particle channel with perfect transmission at `infinite temperature').  

{\bf Conclusion.}---We have examined the structure of charge current fluctuations in a simple classical deterministic model of interacting charged particles. We derived an exact closed-form expression for the MGF in equilibrium at arbitrary background charge density,
encoding the FCS of transferred charge. By performing asymptotic analysis, we deduced a number of remarkable properties:
(I) in the presence of charge bias, fluctuations of the integrated current density on the typical timescale are described by a Gaussian distribution; at vanishing bias we instead discover a distinctly non-Gaussian profile, thereby establishing that the CLT can
be evaded \emph{despite detailed balance};
(II) cumulants $c_{n}(t)$ exhibit, irrespectively of bias, indefinite temporal growth with \emph{distinct} algebraic exponents, implying \emph{divergent} scaled cumulants;
(III) the scaled CGF yields, via the Legendre transform, a bona-fide large-deviation rate function;
(IV) for finite charge bias, the scaled CGF is a non-analytic function of fugacity $\lambda$,
with a discontinuous third derivative at the critical point $\lambda_{c}=0$.

Singular behavior of scaled cumulants, $s_{n}(t) \to \infty$, may be suggestively interpreted as lack of ``sufficiently strong''
temporal clustering associated with a hierarchy of dynamical multipoint current-density correlations
along a temporal seam attached to a fixed point in space, see Eq.~\eqref{eqn:cumulants}.
Such an anomalous structure is a manifestation of enchanced memory effects that are, like other
unconventional transport phenomena such as finite Drude weights \cite{IN_Drude,DS17,transport_review}
and charge superdiffusion \cite{superuniversality,superdiffusion_review,wei2021quantum}, likely inherently tied to stable (quasi)particles that propagate ballistically though the system. This expectation is further corroborated by a
recent numerical study of the lattice Landau--Lifshitz model \cite{Krajnik_breakdown},
widely viewed as an archetypicical completely integrable system \cite{Takhtajan77,Faddeev_book},
which similarly displays divergent scaled cumulants and absence of Gaussianity in the presence of particle-hole symmetry.
Therefore, despite exhibiting diffusive charge dynamics at the level of the dynamical charge correlations,
integrable systems of this sort do not belong to the same universality class as generic (i.e. ergodic) diffusive systems such as, for example,
the SSEP (whose current fluctuations have been computed analytically in \cite{Derrida09}, complying
with the predictions of the MFT \cite{Bertini02,MFT,Derrida11}).

We conclude by pointing out several most pressing issues. Our expectation is that a similar unconventional behavior of
current fluctuations arise in many other classical and quantum solitonic systems and related integrable modes
(including similar superintegrable automata \cite{GP21,Medenjak22}). We specifically have in mind the distinguished conserved
charges associated with manifest discrete or contiuous symmetries \cite{superdiffusion_review}, while other local conservation laws presumably behave in a regular way. Secondly, while non-analytic scaled CGF $F^{[b]}(\lambda)$ is conventionally understood as a precursor of a dynamical phase transition,  we currently lack a more insightful physical interpretation of the encountered third-order critical point.
It is also important to investigate whether there is any degree of universality in the non-Gaussian probability function of the charge current
at the typical scale.  Last but not least, our model offers an opportunity to translate many of these exciting questions into the realm of nonequilibrium physics by either studying evolution of inhomogeneous initial profiles or mesoscopic driven models.

\paragraph*{\bf Acknowledgements.}
We thank  S. Klapp, K. Klobas, A. Kuniba, G. Misguich, V. Popkov and M. \v{Z}nidari\v c for insightful discussions and N. Smith for a useful comment on the manuscript. \v{Z}K acknowledges support of the Milan Lenar\v{c}i\v{c} foundation.
The work has been supported by ERC (European Research Council) Advanced grant 694544-OMNES (TP), by ARRS (Slovenian research agency) research program P1-0402 (\v ZK, EI, TP), and by the SFB910 (project number 163436311) of the DFG (German Research Foundation) (JS).


\bibliography{fluctuations}

\clearpage
\onecolumngrid

\begin{center}
\textbf{{\Large Supplemental Material}}
\end{center}
\begin{center}
\textbf{{\large Exact anomalous current fluctuations in a deterministic interacting model}}
\end{center}

\tableofcontents

\section{Charged hard-core interacting deterministic lattice gas}
\begin{figure}
\centering	
\vspace{-1mm}
\includegraphics[width=\columnwidth]{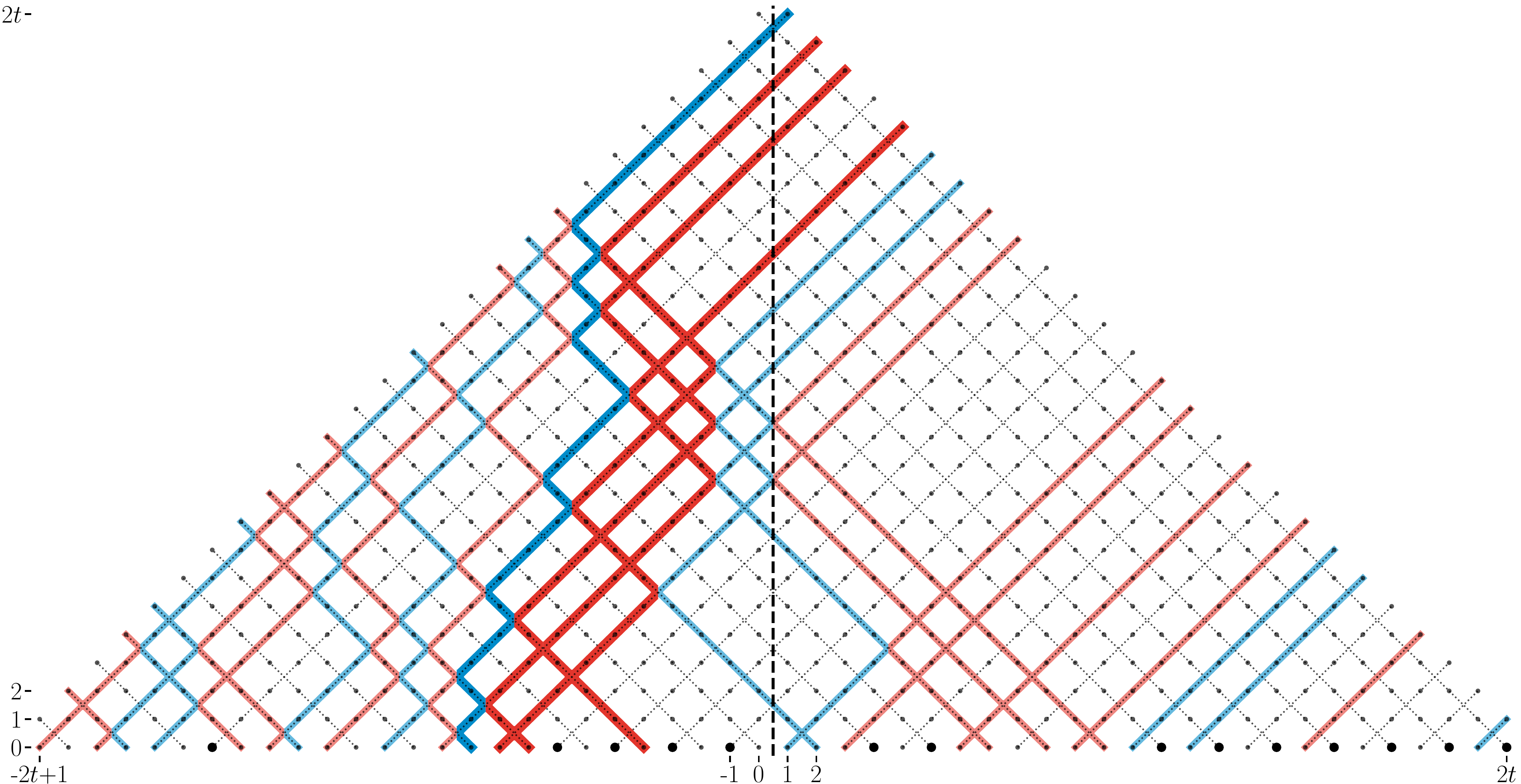}
\vspace{-1mm}
\caption{Coordinate frame (time vertical, space horizontal) of a deterministic charged hardcore lattice gas (red: $+$ particles, blue: $-$ particles, while thin black lines indicate vacancies). Example of a pyramid section of a typical trajectory, for which initial data on a saw of $4t$ subsequent links uniquely determine the transport through the mid-point (dashed line) for 
all times from $0$ to $2t$.}
\label{figHPG1}
\end{figure}

We consider a deterministic lattice gas of charged particles with three local states $\mathcal{Q} = \{\emptyset, +, -\}$, with the corresponding charges $\qr \in \{0, +1, -1\}$ (which we identify with the states from $\mathcal{Q}$), moving on a light-cone lattice. The particles move to the right/left with velocities $\pm 1$ respectively.  Each vertex of the lattice corresponds to a hard-core scattering of two particles according to a deterministic local rule
\begin{equation}
(\emptyset, \emptyset) \leftrightarrow (\emptyset, \emptyset), \qquad (\emptyset, \qr) \leftrightarrow (\qr,\emptyset), \qquad (\qr,\qr') \leftrightarrow (\qr,\qr'), \qquad \textrm{for}\quad \qr,\qr' \in \{+, -\}. \label{two_body_scattering}
\end{equation}
Time evolution of a particle with charge $\qr_\ell^t$ on a space-time lattice $(\ell, t) \in \mathbb{Z}^2$ is generated by staggered applications of the map elementary two-body map $\Phi: \mathcal Q \times \mathcal Q \rightarrow  \mathcal Q \times \mathcal Q$ encoding the local update rule (\ref{two_body_scattering}) 
(see Fig.~\ref{figHPG1} for illustration)
\begin{equation}
(\qr_\ell^{t+1}, \qr_{\ell+1}^{t+1}) = \Phi(\qr_\ell^t, \qr_{\ell+1}^t), \quad \ell+t \equiv 1 \ (\textrm{mod}\ 2).
\end{equation}

\section{Exactly solved full counting statistics}
\subsection{Time integrated current}
\noindent 
We shall consider the total charge which is transferred between the left and right half of the system (through the origin $\ell=0$) in time $2t$ (after $t$ full time steps)
\begin{equation}
\Jr(t) = \sum_{\ell > 0} \qr^{2t}_\ell - \sum_{\ell>0} \qr^0_\ell.
\label{eq:tranQ}
\end{equation}
This is exactly equal to the the time integrated current
\begin{equation}
\Jr(t) = \sum_{t'=0}^{t-1}  \jr_0^{2t'} = \sum_{t'=0}^{2t-1} (-1)^{t'+1} \qr_0^{t'}, \label{J_def}
\end{equation}
where $\jr$ is the local current that satisfies a pair (due to even-odd staggering) of continuity relations 
\begin{equation}
\qr_{2\ell}^{2t+2} - \qr_{2\ell}^{2t} + \jr_{2\ell+1}^{2t+1} - \jr_{2\ell}^{2t} = 0, \qquad \qr_{2\ell+1}^{2t+2} - \qr_{2\ell+1}^{2t} + \jr_{2\ell+2}^{2t} - \jr_{2\ell+1}^{2t+1} = 0.
\end{equation}
A valid expression for the local current is a discrete forward difference, which was used in the second equality in (\ref{J_def})
\begin{equation}
\jr_\ell^t = \qr_\ell^{t+1} - \qr_\ell^t.
\end{equation}
\subsection{Exact moment generating function}
\begin{figure}
\centering	
\vspace{-1mm}
\includegraphics[width=\columnwidth]{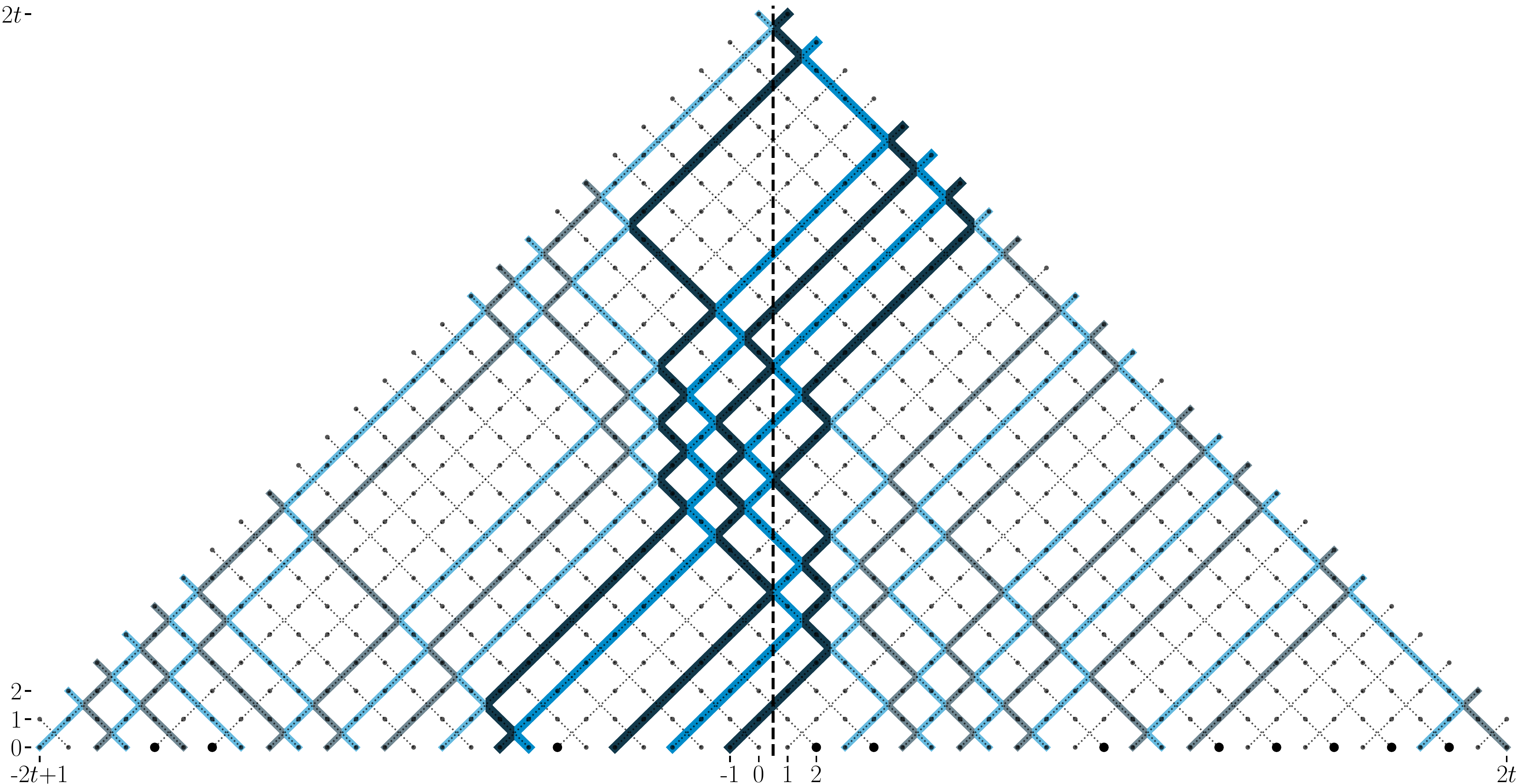}
\vspace{-1mm}
\caption{Distinct particle worldlines within a typical many-body trajectory, depicted with two different colors. Lines crossing the origin are thickened. In this specific example $\Lambda_- = \{-9,-8,-5,-3,-1\}$, $\Lambda_+=\emptyset$,
i.e. there are five worldlines which in total duration $2t=26$ cross the boundary -- indicated by the dashed vertical line -- from left to right sublattice.
}
\label{fig:worldlines}
\end{figure}

Our aim is to compute the moment generating function (MGF)
\begin{equation}
G(\lambda|t) \equiv \langle e^{\lambda \Jr(t)} \rangle \equiv \sum_{\Jr} \mathcal{P}(\Jr|t) e^{\lambda \Jr}, \label{g_def}
\end{equation}
where $\lambda \in \mathbb{C}$ is the counting field and $\langle \bullet \rangle$ denotes the average over an invariant separable measure of initial configurations $\qr_\ell \equiv \qr_\ell^0$
\begin{equation}
\mathbb{P}(\{\qr_{\ell}\})=\prod_{\ell}p(\qr_{\ell}), \qquad p(\pm)=\rho \frac{1 \pm b}{2}, \qquad p(\emptyset)=\ol{\rho}=1-\rho, \label{measure}
\end{equation}
with $0 \leq \rho \leq 1$ the density of particles,  $-1 \leq b \leq 1$ the charge bias of the particles and $\overline \rho$ the density of vacancies. For later convenience we also introduce
\begin{equation}
\varDelta^2 = \rho \overline{\rho} \in [0,1/4].
\end{equation}
The idea is now to evaluate the average (\ref{g_def}) in a {\em nested} way - an outer average over all relevant charge-less (neutral) particle configurations $\{\Sigma\}$, $\Sigma\subset\mathbb Z$, 
and an inner average, with a frozen particle configuration $\Sigma$, over all combinations of
charges $\bigl\{\qr_\ell \in\{+1,-1\}\bigr\}_{\ell \in \Sigma}$. Let us tag a particle starting from the initial configuration and follow its worldline. Note that distinct worldlines by definition
of the dynamics cannot cross. Let $\Lambda_\pm  \subset \Sigma$ denote the coordinates of initial particles, which after
$t$ time-steps (i.e., at time $2t$) end on the opposite side of the lattice, i.e. passing from the interval $[-\infty, 0]$ to $[1, \infty]$ for
$\Lambda_-$ and the opposite for $\Lambda_+$. Note
that due to non-crossing of worldlines, at most one of the subsets $\Lambda_\pm$ can be non-empty (see Fig.~\ref{fig:worldlines} for an illustration).
Considering the defining expression for the transported charge (\ref{eq:tranQ}), an unbiased averaging over the measure (\ref{measure}) given a fixed occupancy configuration $\Sigma$ yields
\begin{equation}
G(\lambda|t;\Sigma) = \prod_{\ell \in  \Lambda_-} \langle e^{\lambda q_\ell} \rangle \prod_{\ell \in  \Lambda_+} \langle e^{-\lambda q_\ell} \rangle = \mu_-^{|\Lambda_-|} \mu_+^{|\Lambda_+|}, \label{sig_g}
\end{equation}
where
\begin{equation}
 \mu_\pm = \cosh \lambda \mp b \sinh \lambda, \label{mu_def}
\end{equation}
and 
$|\Lambda_\pm| \equiv \sum_{\ell \in \Lambda_\pm} 1$. 
Note that the only way
for a worldline to shift left/right is to “meet” a right/left moving vacant site.  It follows that the total signed number of worldlines crossing the origin up to time $2t$ is equal to the number of  right-moving vacancies to the left of the origin at $t=0$, minus the number of left-moving vacancies to the right of the origin again at $t=0$ within the causal cone
\begin{equation}
|\Lambda_+| - |\Lambda_-| = l-r,  \quad l = \sum_{\tau=1}^{t} \delta_{s_{-2\tau+1}, 0}, \quad r = \sum_{\tau=1}^{t} \delta_{s_{2\tau}, 0}.
\end{equation}
Since at most one of the two sets $\Lambda_\pm$ is non-empty we have the identity
\begin{equation}
|\Lambda_+| + |\Lambda_-| = \big||\Lambda_+| - |\Lambda_-|\big|
\end{equation}
allowing us to express
\begin{equation}
|\Lambda_\pm| = \frac{|l-r| \pm (l-r)}{2}.
\end{equation}
By simply counting the number of configurations of the relevant sublattices of vacancies at $t=0$, weighting them with appropriate probabilities in terms of powers of $\rho$ and $\bar{\rho}$, and using the expression (\ref{sig_g}) for each fixed configuration, we obtain the exact moment generating function in the form of a double sum
\begin{gather}
G(\lambda|t) =\rho^{2t} \sum_{l=0}^{t}\sum_{r=0}^{t} \binom{t}{l} \binom{t}{r} \nu^{l+r}  \mu_-^{|\Lambda_-|}\mu_+^{|\Lambda_+|}, \label{g_exact}
\end{gather}
where
\begin{equation}
\nu = \frac{\overline \rho}{\rho}.
\end{equation}
The moment generating function at the origin is normalized to unity
\begin{equation}
G(0|t) = 1, \label{g_norm}
\end{equation}
and is an even function of $\lambda$ as follows from microscopic reversibility of the equilibrium state (\ref{measure})
\begin{equation}
G(\lambda|t) = G(-\lambda|t). \label{even}
\end{equation}
When $|b|=1$ there is only one species of charged particles and the model reduces to that of free ballistically propagating particles, which we refer to as the \emph{free point} of the model.
At the free point the sum (\ref{g_exact}) can be computed exactly
\begin{equation}
G^{[1]}(\lambda|t) = \left[ 1+ 2\varDelta^2(\cosh \lambda -1) \right]^t.
\end{equation}

\subsection{Exact integrated current distribution}
The definition of the moment generating function (\ref{g_def}) can be inverted to obtain the probability distribution of the integrated current by introducing $\chi= e^\lambda$ and using the $Z$-transform (discrete Laplace transform)
\begin{equation}
\mathcal{P}(\Jr|t) = \frac{1}{2\pi \ii} \oint_{|\chi|=1} G(\lambda(\chi)|t) \chi^{-\Jr} \frac{\dd \chi}{\chi}. \label{Z_inversion}
\end{equation}
The normalization of the moment generating function (\ref{g_norm}) induces a normalization of the probability distribution
\begin{equation}
\sum_{\Jr=-\infty}^{\infty} \mathcal P(\Jr|t) = G(0|t) = 1.
\end{equation}
Expanding (for $|b|<1$) the hyperbolic functions in (\ref{g_exact}) with (\ref{mu_def}) using the binomial theorem
\begin{equation}
G(\lambda(\chi)|t) = \rho^{2t} \sum_{l,r= 0}^{t} \binom{t}{r} \binom{t}{l} \nu^{l+r} \sum_{i_\pm=0}^{|\Lambda_\pm|} \binom{|\Lambda_-|}{i_-} \binom{|\Lambda_+|}{i_+}
 \left[ \frac{1-b}{2} \right]^{|\Lambda_-| - i_- + i_+} \left[ \frac{1+b}{2} \right]^{|\Lambda_+|-i_++i_- }  \chi^{2(i_++i_-) -|l-r|}, \label{G_binom}
\end{equation}
 and computing the integral (\ref{Z_inversion}) by the residue theorem we find the exact current distribution
\begin{equation}
\mathcal P (\Jr|t) = \rho^{2t}\sum_{l,r=0}^{t} \binom{t}{r} \binom{t}{l} \binom{|l-r|}{\frac{|l-r|+\Jr}{2}} \nu^{l+r} \left[\frac{1-b^2}{4}\right]^{\frac{|l-r|}{2}}\left[\frac{1-b}{1+b}\right]^{ \frac{\Jr\, \textrm{sgn}\, (l-r)}{2}} 
\frac{1+(-1)^{\Jr + |l-r|}}{2}, \label{P_exact}
\end{equation}
where we adopt the convention
\begin{equation}
\textrm{sgn}\, (x) =
\begin{cases}
\hfill +1 \hfill & \textrm{for } x >0,\\
\hfill \ \ 0  \hfill& \textrm{for } x = 0,\\
\hfill -1 \hfill &  \textrm{for } x<0.
\end{cases}
\end{equation}
The free points $ b=\pm 1$ have to be considered separately and can be summed up using a hypergeometric function
\begin{equation}
\mathcal P^{[1]}(\Jr|t) =  \nu^{\mp \Jr} \rho^{2t} \sum_{l=0}^t \binom{t}{l} \binom{t}{l \mp \Jr} \nu^{2l} = \nu^{\Jr}  \rho^{2t} \binom{t}{\Jr} {}_2F_1\left(\Jr -t, -t; \Jr+1; \nu^{2}\right).
\end{equation}

\section{Asymptotic analysis}
\subsection{Preliminaries}
In this section we analyse the late-time asymptotics of the exact results (\ref{g_exact},\ref{P_exact}) and various aspects thereof. We emphasize that throughout this section we specialize our analysis to the case of a real counting field
\begin{equation}
\lambda \in \mathbb{R}.
\end{equation}
The free point of the model has already been considered, so we restrict ourselves to  the interacting regime
\begin{equation}
-1 < b <1.
\end{equation}
Throughout the paper and supplemental material we use the following asymptotic notation
\begin{align}
f \stackrel{x}{\sim} g &\Leftrightarrow  \lim_{x \to \infty} \frac{f(x)}{g(x)} = c \in \mathbb{C} \setminus  \{0\},\\
f \stackrel{x}{\asymp} g &\Leftrightarrow  \lim_{x \to \infty} \frac{f(x)}{g(x)} = 1,
\end{align}
where the variable above the asymptotic relation indicates which variables is sent to infinity. We most frequently send $t \to \infty$, in which case we suppress the variable above the asymptotic relation.\\\\
 In what follows we frequently make use of Laplace's method \cite{miller2006applied} of localizing exponential integrals involving a large parameter over multi-dimensional domains spanned by coordinates $\vec{x} = (x_1, x_2, \ldots, x_n)$ as 
\begin{equation}
 \int \dd \vec{x}\ g(\vec{x}) e^{t f(\vec{x})} \asymp  \left(\frac{2\pi}{t}\right)^{n/2} \frac{g(\vec{x}_0)e^{tf(\vec{x}_0)}}{\sqrt{|\det[H[f](\vec x_0)]|}}, \qquad \nabla f(\vec x_0) = \vec 0, \label{Laplace}
\end{equation}
where $H[f]$ is the Hessian matrix of function $f$. We also make use of the Stirling formula
\begin{equation}
n! \stackrel{n}{\asymp} \sqrt{2\pi n} \left( \frac{n}{e} \right)^n, \label{Stirling}
\end{equation}
in particular to approximate binomials as
\begin{equation}
\binom{t}{l}  \stackrel{t-l,t, l}{\asymp} \sqrt{\frac{t}{2\pi l(t-l)}}\left(\frac{l}{t}\right)^{-l}  \left(1-\frac{l}{t}\right)^{l-t}, \label{binom_approx}
\end{equation}
where the order of limits, implied by the asymptotic notation, can be permuted.

\subsection{Moment generating function}
The dominant terms of the sum (\ref{g_exact}) are located away from the boundaries $l, r = 0$ and $l, r = t$.
It is then justified to  approximate the binomials using (\ref{binom_approx}), convert the sums into integrals and introduce rescaled variables $x = l/t$, $y=r/t$
\begin{equation}
G(\lambda|t) \asymp \frac{t}{2\pi} \iint_{[0, 1]^2}  \frac{e^{tf(x,y)}}{\sqrt{xy(1-x)(1-y)}}\, \dd x \dd y,  \label{g_int}
\end{equation}
where
\begin{equation}
f(x, y) =\log \left( \rho^2 \frac{\nu^{x+y} \mu_+^{\frac{1}{2}(|x-y|+(x-y))} \mu_-^{\frac{1}{2}(|x-y|-(x-y))} }{x^x (1-x)^{1-x}y^y (1-y)^{1-y}}\right). \label{saddle_f}
\end{equation}
As $t \to \infty$ the integral localizes in the vicinity of the two extrema of the function in the exponent
\begin{gather}
\nabla f(x_\pm, y_\pm) = (0,0), \quad x_\pm= \frac{\nu}{\nu + \mu_\pm^{\mp1}}, \quad y_\pm = \frac{\nu}{\nu + \mu_\pm^{\pm 1}} \label{max1}.
\end{gather}
By inspecting the values at the these two points we observe that the global  maximum is attained at
\begin{equation}
\max_{(x, y) \in [0,1]^2} f(x, y) = f(x_\gamma, y_\gamma), \quad \gamma = -\textrm{sgn}\,(\lambda b), \label{max_func}
\end{equation}
so that generically there is only one global maximum. The generic picture breaks down when $\lambda = 0$ or $b=0$. For $\lambda = 0$ and $b \neq 0$ we have $x_+ = x_- = y_+ = y_-$ so that there is still a single global maximum on the diagonal of the unit square $[0,1]^2$. The case of $b = 0$, $\lambda \neq  0$ is more complicated since there the function $f$ attains its global maximum at two distinct points, symmetric with respect to the  diagonal, $x=y$. This introduces a discontinuity in the function $G$ at $\lambda = 0$.
Evaluating the required objects at the maximum, namely
\begin{gather}
\det H(x_\gamma, y_\gamma) =  \left((\mu_\gamma + \nu)(\mu_\gamma^{-1} + \nu)\nu^{-1}\right)^{2},\\
  \frac{1}{\sqrt{x_\gamma(1-x_\gamma)y_\gamma(1-y_\gamma)}} = (\mu_\gamma + \nu)(\mu_\gamma^{-1} + \nu)\nu^{-1},
\end{gather}
and taking into account the localization of the integral (\ref{g_int}), we find
\begin{equation}
G(\lambda|t) \asymp e^{tf(x_\gamma, y_\gamma)} = D_b(\lambda) \left[\rho^2 (\nu + \mu_\gamma)(\nu + \mu_\gamma^{-1}) \right]^t, \label{g_res}
\end{equation}
where the factor $D$ accounts for the discontinuity at  the origin when $b=0$
\begin{equation}
D_b(\lambda) =
\begin{cases}
2 & \textrm{for }\lambda = b = 0,\\
1 & \textrm{otherwise}.
\end{cases}
\end{equation}

\subsection{Scaled cumulant generating function}
The scaled cumulant generating function (SCGF) is defined as
\begin{equation}
F(\lambda) = \lim_{t \to \infty} \frac{1}{t}  \log G(\lambda|t). \label{F_def}
\end{equation}
From the the result (\ref{g_res}) it immediately follows that
\begin{equation}
F(\lambda) = \log \left[1 + \varDelta^2 (\mu_{b} + \mu^{-1}_{b} - 2) \right] , \quad \mu_b= \cosh \lambda + |b|\sinh |\lambda|. \label{F_res}
\end{equation}
The SCGF has the following expansion around $\lambda = 0$
\begin{gather}
F^{[b]}(\lambda) = \lambda^2 \varDelta^2 b^2 + |\lambda|^3 \varDelta^2 |b|(1-b^2) + \lambda^4 \frac{\varDelta^2}{12}\left(3  - 14b^2 + 12b^4 - 6 \varDelta^2 b^4 \right)   + \mathcal{O}(|\lambda|^5), \label{Fn0}\\
F^{[0]}(\lambda) = \lambda^4 \frac{\varDelta^2}{4} - \lambda^6 \frac{\varDelta^2}{12} + \lambda^8  \frac{\varDelta^2 (11 -10 \varDelta^2)}{320} +  \mathcal{O}(\lambda^{10}), \label{F0}
\end{gather}
and around $|\lambda| \to \infty$
\begin{gather}
F(\lambda) = |\lambda| + \log \left(\frac{\varDelta^2(1+|b|)}{2} \right) + \frac{2(1-2\varDelta^2)}{\varDelta^2 (1+|b|)}e^{-|\lambda|}+ \mathcal{O}(e^{-2|\lambda|}). \label{inf_asymp}
\end{gather}
We note that while the asymptotic MGF (\ref{g_res}) for $b=0$ is discontinuous at the origin, the SCGF (\ref{F_res}) is in this case a real analytic function.
On the contrary the SCGF for $b\neq 0$ has a discontinuous third derivative at the origin
\begin{equation}
\frac{\dd}{\dd \lambda}\Big|_{\lambda = 0} F^{[b]}(\lambda)  = 0, \qquad
\frac{\dd^2}{\dd \lambda^2}\Big|_{\lambda = 0} F^{[b]}(\lambda)  = 2b^2 \varDelta^2, \qquad
\frac{\dd^3}{\dd \lambda^3}\Big|_{\lambda = 0^\pm} F^{[b]}(\lambda)  = \pm 6|b|(1-b^2) \varDelta^2, \label{broken_der}
\end{equation}
where $0^\pm$ indicates from which side we approach the origin.

\subsection{Cumulants}
The cumulant generating function (CGF) is defined as the logarithm of the moment generating function (\ref{g_def})
\begin{equation}
 \log G(\lambda|t). \label{CGF_def}
\end{equation}
We define the raw cumulants as derivatives of the cumulant generating function at the origin
\begin{equation}
c_n(t) \equiv \frac{\dd^n}{\dd \lambda^n} \Big|_{\lambda = 0} \log G(\lambda|t). \label{cum_def}
\end{equation}
In equilibrium, the CGF is an even function [see Eq.~(\ref{even})], implying that all odd cumulants identically vanish
\begin{equation}
c_{2n-1}(t) = 0. \label{odd_zero}
\end{equation}
We note that the existence of a non-zero SCGF (\ref{F_def}) does not imply that at late times the cumulants all grow lineary with time since the limit $t \to \infty$ and the operation of taking derivatives with respect to $\lambda$  in general do not commute
\begin{equation}
\frac{\dd^n}{\dd \lambda^n} \Big|_{\lambda= 0  } \lim_{t \to \infty} \frac{1}{t}\log G(\lambda|t) \neq  \lim_{t \to \infty} \frac{\dd^n}{\dd \lambda^n}\Big|_{\lambda= 0  } \frac{1}{t}  \log G(\lambda |t). \label{non_commute}
\end{equation}
The discontinuity of the SCGF (\ref{broken_der}) already indicates that cumulants cannot be obtained by taking derivatives of the SCGF, but need to be computed directly from (\ref{cum_def}). The combinatorial sums inherent in the definition of the cumulants (\ref{cum_def}) can be  resolved using the Fa\`{a} di Bruno's formula for higher derivatives of a composite function
\begin{equation}
\frac{\dd^n}{\dd x^n} f(g(x)) = \sum_{r=1}^n f^{(r)}(g(x)) B_{n, r}\left(g^{(1)}(x),  g^{(2)}(x), \ldots, g^{(n-r+1)}(x) \right), \label{FdB}
\end{equation}
where $f^{(k)} \equiv \frac{\dd^k f}{\dd x^k}$ and $B_{n, r}$ are the incomplete Bell polynomials, defined by
\begin{equation}
 \exp \left[{z \sum_{j=1}^\infty x_j \frac{t^j}{j!}}\right]=1+ \sum_{n=1}^\infty \frac{t^n}{n!}\sum_{k=1}^n z^k B_{n, k}(x_1, x_2, \ldots, x_{n-k+1}).
\end{equation}

\subsubsection{Localization of cumulants}
Consider first the computation of the $n$-th moment $G^{(n)}$ by taking derivatives of the MGF. Start by defining functions $d_n^k$ as
\begin{equation}
d_{n}^k(b) \equiv \frac{\dd^n}{\dd \lambda^n} \Big|_{\lambda = 0} \left(\cosh \lambda + b \sinh \lambda \right)^{k} = \sum_{r=1}^n \frac{k!}{(k-r)!} B_{n, r}\left( b, 1,b, 1, \ldots \right) \label{Faa_cn}
\end{equation}
so that the $n$-th moment can be expressed as
\begin{equation}
G^{(n)}(t) \equiv \frac{\dd^n}{\dd \lambda^n} \Big|_{\lambda = 0}  G(\lambda|t) = \rho^{2t}\sum_{l=0}^{t}\sum_{r=0}^{t} \binom{t}{l} \binom{t}{r} \nu^{l+r} d_n^{|l-r|} (-b\,  \textrm{sgn}(l-r)). \label{Faa_gn}
\end{equation}
All odd functions $d_{2n+1}^k(b)$ contain only odd powers of $b$, while all even functions $d_{2n}^k(b)$ contain only even powers of $b$.
From the symmetric way in which $b$ enters $d_n^k$ in Eq.~(\ref{Faa_gn}) we see that all odd moments are zero
\begin{equation}
G^{(2n-1)}(t) = 0, \label{odd_moments}
\end{equation}
while the even terms do not depend on the sign of $b$.
To evaluate Eq.~(\ref{Faa_gn})  we consider sums of the form
\begin{equation}
S_n = \rho^{2t}\sum_{l=0}^{t}\sum_{r=0}^{t} \binom{t}{l} \binom{t}{r} \nu^{l+r} |l-r|^n. \label{S_def}
\end{equation}
Approximating the binomials with (\ref{binom_approx}), converting the sum into an integral and rescaling the integral (\ref{S_def}) as when computing the SCGF ($x=l/t$, $y=r/t$), we obtain
\begin{equation}
S_n \asymp \frac{t^{n+1}}{2\pi} \iint_{[0, 1]^2}  h_n(x, y) e^{tf(x,y)}\, \dd x \dd y, 
\end{equation}
where 
\begin{equation}
f(x, y) =\log \left( \rho^2 \frac{\nu^{x+y}}{x^x (1-x)^{1-x}y^y (1-y)^{1-y}}\right), \quad h_n(x, y) = \frac{|x-y|^n}{\sqrt{xy(1-x)(1-y)}}.
\end{equation}
The gradient of $f$ vanishes at the critical point
\begin{equation}
\nabla f(x_c, y_c) = 0, \quad x_c = y_c = \frac{\nu}{1 + \nu} = 1 - \rho,
\end{equation}
with the Hessian and value
\begin{equation}
H(x_c, y_c) = - \mathds{1}/\varDelta^2, \quad \det H(x_c, y_c) = \varDelta^{-4}, \quad f(x_c, y_c) = 0.
\end{equation}
Note however, that $h_n(x_c, y_c) = 0$ for all $n > 0$ and simple localization (\ref{Laplace}) fails since there is a thin strip where $h$ is small, which splits the peak into two symmetric peaks. To capture this behavior we Taylor expand $f$ around the peak to quadratic order
\begin{equation}
f(x, y) \approx -\frac{(x - (1-\rho))^2 + (y-(1-\rho))^2}{2\varDelta^2},
\end{equation}
and introduce rotated coordinates
\begin{equation}
s_\pm = x \pm y, \label{rot_coord} 
\end{equation}
in terms of which the critical point is at
\begin{equation}
 s_-^c = 0, \quad s_+^c = 2(1-\rho).
\end{equation}
Given that the two peaks are symmetric around $s_-=0$ we compute twice the integral over the peak with $s_- > 0$
\begin{equation}
S_n \asymp \frac{t^{n+1}}{\pi} \iint_{s_- > 0} \frac{ s_-^n e^{-\frac{t}{4\varDelta^2}(s_-^2)}  e^{-\frac{t}{4\varDelta^2}(s_+ - s_+^c)^2} }{\sqrt{(s_+^2 - s_-^2)(1 - \frac{s_+ +s_-}{2})(1 - \frac{s_+ -s_-}{2})}} \dd s_- \dd s_+,
\end{equation}
The expression under the square root is not singular at the peak and localizes to $\frac{1}{\varDelta^2}$
\begin{equation}
S_n \asymp \frac{t^{n+1}}{2\pi \varDelta^2} \int_{-\infty}^{\infty} \dd s_+ e^{-\frac{t}{4\varDelta^2} (s_+ -s_+^c)^2} \int_{0}^\infty \dd s_- s_-^n e^{-\frac{t}{4\varDelta^2} s_-^2}.
\end{equation}
Computing the Gaussian integrals we obtain the late-time behavior of the sum (\ref{S_def})
\begin{equation}
S_n \asymp t^{n/2} \frac{(2\varDelta)^{n} }{\sqrt{\pi}}\Gamma \left(\frac{n+1}{2}\right). \label{S_res}
\end{equation}
Having obtained the result (\ref{S_res}) we are in a position to extract the asymptotic behavior of the moments and cumulants.
 Equation~(\ref{Faa_cn}) generates $d_n^k$ as a function of $k$. Expanding, we obtain a polynomial in $k$ of degree $n/2$ . Substituting the monomials as [cf. Eqs.~(\ref{Faa_gn},\ref{S_def})]
\begin{equation}
k^m \to S_m,
\end{equation}
we come to the expression for the even moments $G^{(2n)}$ in terms of  $S_m$, for which we have the asymptotic result (\ref{S_res})
\begin{equation}
G^{(2n)}(t) \asymp \rho^{2t}\sum_{l=0}^{t}\sum_{r=0}^{t} \binom{t}{l} \binom{t}{r} \nu^{l+r} d_{2n}^{|l-r|} (b) \xrightarrow{k^m \to S_m} G^{(2n)}(S_1, S_2, \ldots). \label{Faa_final}
\end{equation}
To pass from the moments to the cumulants we again invoke (\ref{FdB}) while taking higher derivatives of the logarithm
\begin{equation}
c_n(t) \asymp   \sum_{r=1}^n (-1)^{r-1} (r-1)! B_{n, r}\left(0, G^{(2)}(t), 0, G^{(4)}(t), \ldots, G^{(n-r+1)}(t) \right). \label{Faa_log}
\end{equation}
where we make use of the fact that all odd moments are zero (\ref{odd_moments}).

\subsubsection{Lowest cumulants}
\begin{figure}[h!]
\centering	
\includegraphics[width=0.55\columnwidth]{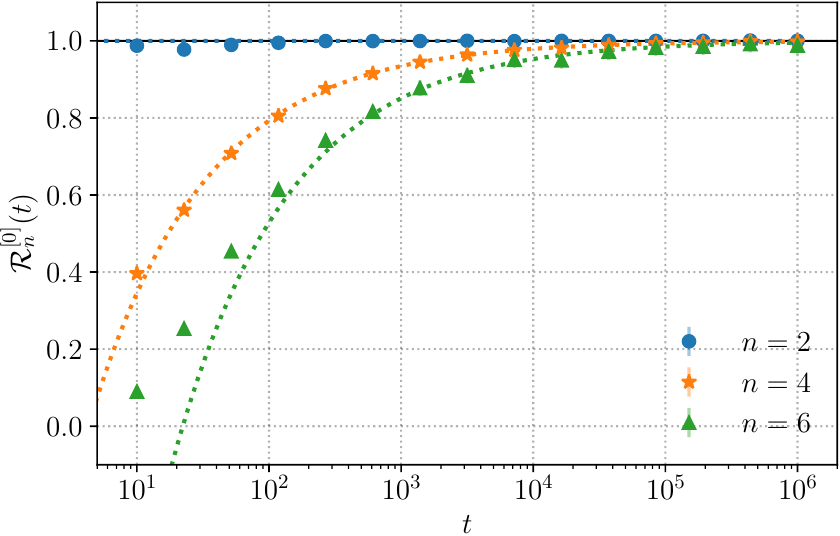}
\caption{Ratio  $\mathcal{R}^{[0]}_n$ (\ref{ratio_def})  (circles/stars/triangles) of numerically estimated cumulants $\tilde c_n(t)$ divided by leading order asymptotics $c_{n|0}^{[0]} t^{n/4}$ as a function of time $t$ for $\rho=0.5$ ($\varDelta=0.25$) and $b=0$. Dotted lines show the asymptotic results including an additional sub-leading order,  $1 + t^{-1/2} c_{n|1}^{[0]}/{c_{n|0}^{[0]}} $. Number of samples for each time ${\rm N}_{\rm sample} = 10^{8}$.}
\label{fig_cum1}
\end{figure}

\begin{figure}[h!]
\centering	
\includegraphics[width=0.55\columnwidth]{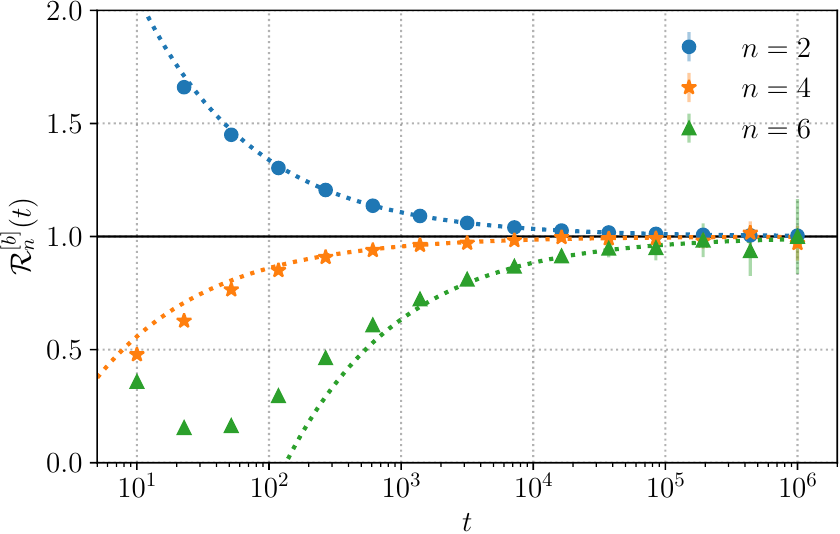}
\caption{Ratio $\mathcal{R}^{[b]}_n$ (\ref{ratio_def}) (circles/stars/triangles) of numerically estimated cumulants $\tilde c_n(t)$ divided by leading order asymptotics $c_{n|1}^{[0]} t^{n/2}$ (or $c_{n|0}^{[0]} t^{n/2}$ for $n=2$) as a function of time $t$ for $\rho=0.5$ ($\varDelta=0.25$) and $b=0.5$. Dotted lines show the asymptotic results including an additional sub-leading order,  $1 + t^{-1/2}  c_{n|2}^{[0]}/c_{n|1}^{[0]}$ (or $1 +t^{-1/2}  c_{n|1}^{[0]}/c_{n|0}^{[0]} $ for $n=2$). Number of samples for each time ${\rm N}_{\rm sample} = 10^{8}$.}
\label{fig_cum2}
\end{figure}

Combining the results (\ref{S_res},\ref{Faa_final},\ref{Faa_log}) it is straightforward to extract the late-time behavior of cumulants order-by-order
\begin{align}
c_n^{[0]}(t) &= \sum_{l=0}^r c_{n|l}^{[0]} t^{(n-2l)/4} + \mathcal{O}(t^{(n-2(r+1))/4}), \label{asymp_series_b0}\\
c_n^{[b]}(t) &= \sum_{l=0}^r c_{n|l}^{[b]} t^{(n-l)/2} + \mathcal{O}(t^{(n-(r+1))/2}). \label{asymp_series_b1}
\end{align}
As an example we list the first few leading orders of the lowest cumulants. For $b=0$ we have
\begin{align}
c_{2|0}^{[0]} &= \frac{2\varDelta}{\pi^{1/2}}, \label{ref_c2}\\
c_{4|0}^{[0]} &= \frac{6\varDelta^2}{\pi} \left(\pi -2 \right), &
c_{4|1}^{[0]}&=  - \frac{4\varDelta}{\pi^{1/2}}, \label{ref_c4}\\
c_{6|0}^{[0]}&= \frac{60 \varDelta^3}{\pi^{3/2}}(4 - \pi), &
c_{6|1}^{[0]}  &=  \frac{60 \varDelta^2}{\pi}(2-\pi), \label{ref_c6}\\
c_{8|0}^{[0]} &=  \frac{3360 \varDelta^4}{\pi^2}(\pi - 3),  &
c_{8|1}^{[0]}&=  - \frac{1680 \varDelta^3}{\pi^{3/2}} \left(4 - \pi\right) , \label{ref_c8}\\
c_{10|0}^{[0]}&= \frac{7560\varDelta^5}{\pi^{5/2}}(3\pi^2 - 40 \pi + 96), & 
c_{10|1}^{[0]} &=  \frac{201600\varDelta^4}{\pi^2}\left(3 - \pi \right). \label{ref_c10}
\end{align}
while for $0<|b|<1$ we find
\begin{align}
c_{2|0}^{[b]} &= 2 \varDelta^2 b^2 & 
c_{n>2|0}^{[b]} &= 0 \label{ref_cb0}\\
c_{2|1}^{[b]} &= \frac{2\varDelta}{\sqrt{\pi}}(1-b^2),\label{ref_cb2}\\
c_{4|1}^{[b]} &= \frac{24}{\sqrt{\pi}} \varDelta^3 b^2(1-b^2)  &
c_{4|2}^{[b]}  &= \frac{6\varDelta^2(1-b^2)}{\pi}\left[(1-b^2)(\pi-2) - \frac{8}{3}\pi b^2\right], \label{ref_cb4}\\
c_{6|1}^{[b]} &= -\frac{120}{\sqrt{\pi}} \varDelta^5 b^4(1-b^2) &
c_{6|2}^{[b]}  &= \frac{360b^2 \varDelta^4}{\pi}  \left(\pi - 2\right) (1-b^2)^2, \label{ref_cb6}\\
c_{8|1}^{[b]} &= \frac{1344}{\sqrt{\pi}} \varDelta^7 b^6(1-b^2) & 
c_{8|2}^{[b]} &=   -  \frac{13440\varDelta^6 b^4}{\pi}\left(1-b^2\right)^2, \label{ref_cb8}\\
c_{10|1}^{[b]}  &=   -\frac{21600}{\sqrt{\pi}} \varDelta^9 b^8(1-b^2) &
c_{10|2}^{[b]} &=   \frac{483840 \varDelta^8 b^6}{\pi}(1-b^2)^2. \label{ref_cb10}
\end{align}
In Figures~\ref{fig_cum1},\ref{fig_cum2} we show the comparison of asymptotics (\ref{ref_c2},\ref{ref_c4},\ref{ref_c6},\ref{ref_cb0},\ref{ref_cb2},\ref{ref_cb4},\ref{ref_cb6}) with direct numerical simulations for $b=0$ and $b=0.5$ respectively. In both cases we plot the ratio $\mathcal{R}_n$ of  numerically estimated cumulants $\tilde c_n$ divided by the first respective non-zero asymptotic order (see Eqs.~(\ref{asymp_series_b0},\ref{asymp_series_b1})) 
\begin{equation}
\mathcal{R}^{[0]}_n(t) \equiv
\frac{\tilde c_n(t)}{c_{n|0}^{[0]}}t^{-n/4},\\
 \qquad
\mathcal{R}^{[b]}_n(t) \equiv
\begin{cases}
\frac{\tilde c_n(t)}{c_{n|0}^{[b]}}t^{-1}, & \textrm{for}\ n=2,\\
\frac{\tilde c_n(t)}{c_{n|1}^{[b]}}t^{-(n-1)/2}, & \textrm{for}\  n>2.
\end{cases} 
\label{ratio_def}
\end{equation}

The ratios converge towards unity for large times for both values of charge bias $b$, $ \lim_{t \to \infty} \mathcal{R}_n(t) \to 1$.
We observe that the cumulants scale with different powers of $t$ as
\begin{equation}
c_2^{[b]}(t) \sim t, \quad c^{[b]}_{2n>2}(t) \sim t^{n-1/2} \qquad {\rm and} \qquad c^{[0]}_{2n}(t) \sim t^{n/2}.
\label{cumulant_growth}
\end{equation}

\clearpage
\subsubsection{Generator of cumulant asymptotics for $b=0$}
\noindent To prove the assertion (\ref{cumulant_growth}) and obtain a convenient way of generating the leading coefficients of the expansion (\ref{asymp_series_b0}) we again make use of the Fa\`{a} di Bruno's formula (\ref{FdB}).
We start by considering the case of $b=0$ where the moments are to leading order in time equal to
\begin{equation}
G^{(2n)}(t) \asymp (2n-1)!! S_{n}(t) \asymp (2n-1)!! \frac{(2\varDelta)^n }{\sqrt{\pi}} \Gamma\left( \frac{n+1}{2}\right) t^{n/2}.
\end{equation}
From Eq.~(\ref{Faa_log}) the cumulants are expressed as
\begin{equation}
c_n(t)  \asymp  \sum_{r=1}^n (-1)^{r-1} (r-1)! B_{n, r}\left(0, 1!! \frac{2\varDelta\Gamma\left(1\right)}{\sqrt{\pi}}t^{1/2} , 0,3!! \frac{(2\varDelta)^2\Gamma\left(3/2\right)}{\sqrt{\pi}}t, \ldots \right). \label{Faa_b0}
\end{equation}
To resum the expression we can use (\ref{FdB}).
 We construct a Taylor series from the prescribed derivatives and resum the series for an auxiliary  function $h$
\begin{equation}
h(\lambda, t) \equiv \pi^{-1/2}  \sum_{j=0}^\infty \frac{(2j-1)!!}{(2j)!} \Gamma\left(\frac{j+1}{2}\right)  (2\varDelta)^j t^{j/2}\lambda^{2j}  = e^{\frac{\lambda^4 \varDelta^2 t}{4}} \left(1  + \textrm{erf} \left[ \frac{\lambda^2 \varDelta \sqrt{t}}{2} \right] \right).
\end{equation}
Introducing a rescaled variable $\xi$
\begin{equation}
\xi^2 = \frac{\lambda^2 \varDelta\, t^{1/2}}{2},
\end{equation}
and noting that $h(0, t) = 1$ we can apply the Fa\`{a} di Bruno formula (\ref{FdB}) in reverse to resum the expression (\ref{Faa_b0})
\begin{equation}
c_n(t) \asymp \frac{\dd^n}{\dd \lambda^n} \Big|_{\lambda = 0}  \mathcal F^{[0]}_0(\xi(\lambda, t)) = \left( \frac{\dd \xi}{\dd \lambda} \right)^n \frac{\dd^n}{\dd \xi^n} \Big|_{\xi = 0} \mathcal F^{[0]}_0(\xi),  \label{Faa_b0_0}
\end{equation}
where
\begin{equation}
 \mathcal F^{[0]}_0(\xi) = \xi^4 + \log \left(1 + \textrm{erf}\ \xi^2 \right). \label{Faa_b0_1}
\end{equation}
The expression (\ref{Faa_b0_0}) recovers the leading asymptotics  (\ref{ref_c2},\ref{ref_c4},\ref{ref_c6},\ref{ref_c8},\ref{ref_c10}).
\begin{equation}
c_{n|0}^{[0]} =  \left(\frac{\varDelta}{2}\right)^{n/2} \frac{\dd^n}{\dd \xi^n} \Big|_{\xi = 0} \mathcal F^{[0]}_0(\xi). \label{leading_generator_b0}
\end{equation}
From (\ref{Faa_b0_0},\ref{Faa_b0_1},\ref{leading_generator_b0}) it follows immediately that $c_{2n}^{[0]}(t) \sim t^{n/2}$.

\subsubsection{Generator of cumulant asymptotics for $0<|b|<1$}
A similar procedure can be applied when $0< |b|<1$, with the proviso that the leading order contribution cancels (for $n>2$) and the next sub-leading contribution  of the moments must be included
\begin{align}
G^{(2n)}(t) &\asymp  b^{2n} S_{2n}(t) +  n(2n-1) b^{2n-2}(1-b^2) S_{2n-1}(t),\\
& \asymp \frac{(2 b \varDelta t^{1/2})^{2n-1}}{\sqrt{\pi}} \left[ 2b \varDelta \Gamma (n+\tfrac{1}{2}) t^{1/2} +  n(2n-1) (b^{-1}-b) \Gamma(n) \right].
\end{align}
We yet again construct the Taylor series, resum it and introduce a rescaled variable $\xi$
\begin{equation}
\xi = \lambda b \varDelta t^{1/2},
\end{equation}
 yielding for $t\to \infty$
\begin{align}
h(\lambda, t) &\equiv \frac{1}{\sqrt{\pi}} \sum_{j=0}^\infty \frac{\left[ 2b \varDelta \Gamma (n+\tfrac{1}{2}) t^{1/2} +  n(2n-1) (b^{-1}-b) \Gamma(n) \right]}{(2j)!}  (2b \varDelta t^{1/2})^{2j-1}  \lambda^{2j} \\
 &= h_0(\xi) + t^{-1/2} \,  h_1(\xi) + \mathcal{O}(t^{-1})
\end{align}
where
\begin{equation}
h_0(\xi) =  e^{\xi^2}, \quad  h_1(\xi) = \frac{1-b^2}{b^2 \varDelta} \left(\xi^3 e^{\xi^2} \textrm{erf}\ \xi + \frac{\xi^2}{\sqrt{\pi}} \right). \label{h_eqs}
\end{equation}
We now expand the logarithm for large $t$ with $\xi$ held fixed 
\begin{equation}
c_n(t) = \frac{\dd^n}{\dd \lambda^n} \Big |_{\lambda= 0} \log \left(h_0(\xi) + t^{-1/2}h_1(\xi)  + \mathcal{O}(t^{-1}) \right) = \frac{\dd^n}{\dd \lambda^n} \Big |_{\lambda= 0} \log\left( h_0(\xi) + t^{-1/2} \frac{h_1(\xi)}{h_0(\xi)} + \mathcal{O}(t^{-1}) \right).
\end{equation}
Using the expressions (\ref{h_eqs}) we come to
\begin{equation}
c_n(t) \asymp \frac{\dd^n}{\dd \lambda^n} \Big |_{\lambda= 0} \mathcal F^{[b]}(\xi(\lambda, t))=\left( \frac{\dd \xi}{\dd \lambda} \right)^n \frac{\dd^n}{\dd \xi^n} \Big |_{\xi= 0} \mathcal F^{[b]}(\xi), \label{Faa_b_0}
\end{equation}
where
\begin{equation}
 \mathcal F^{[b]}(\xi) =  \mathcal F_0^{[b]}(\xi)  +\mathcal F_1^{[b]}(\xi)   t^{-1/2} \label{Faa_b_1},
\end{equation}
with
\begin{equation}
\mathcal F^{[b]}_0(\xi) = \xi^2 , \qquad  \mathcal F_1^{[b]}(\xi) = \frac{1-b^2}{b^2 \varDelta} \left[\xi^3 \textrm{erf}\  \xi + \frac{\xi^2e^{-\xi^2}}{\sqrt{\pi}} \right].
\end{equation}
The expression (\ref{Faa_b_0}) recovers the leading asymptotics (\ref{ref_cb0},\ref{ref_cb2},\ref{ref_cb4},\ref{ref_cb6},\ref{ref_cb8},\ref{ref_cb10})
\begin{equation}
c_{n|0}^{[b]} = \left(|b|\varDelta \right)^{n} \frac{\dd^n}{\dd \xi^n} \Big|_{\xi = 0}\mathcal F_0^{[0]}(\xi), \qquad c_{n|1}^{[b]} = \left(|b|\varDelta \right)^{n} \frac{\dd^n}{\dd \xi^n} \Big|_{\xi = 0} \mathcal F_1^{[0]}(\xi). \label{leading_generator_b1}
\end{equation}
From (\ref{Faa_b_0},\ref{Faa_b_1},\ref{leading_generator_b1}) it follows that $c_2^{[b]}(t) \sim t$, $c_{2n>2}^{[b]}(t) \sim t^{n-1/2}$.

\subsubsection{Generator of cumulant asymptotics within the Gaussian approximation}
A single generator of leading cumulant asymptotics, valid for all $|b|<1$, can alternatively be obtained by starting from the exact current distribution (\ref{P_exact}) and using the De Moivre-Laplace theorem to approximate the binomial
\begin{equation}
\binom{t}{l}p^l q^{t-l} \asymp \frac{1}{\sqrt{2\pi t pq}}\exp\left[ - \frac{(l - tp)^2}{2tpq} \right] \quad{\rm for}\ l - tp \sim t^{1/2} \ {\rm and }\ p+ q = 1. \label{Gaussian_approx}
\end{equation}
Note that the approximation (\ref{Gaussian_approx}) is less precise than the Stirling approximation (\ref{binom_approx}).
We start by splitting the probability distribution as
\begin{equation}
\mathcal{P}(\Jr|t) = \sum_{l-r}\mathcal{P}_1(\Jr|l-r) \mathcal{P}_2(l-r|t), \label{p_split}
\end{equation}
where
\begin{align}
\mathcal{P}_1(\Jr|l-r)  &\equiv \binom{|l-r|}{\frac{|l-r|+\Jr}{2}} \left[\frac{1-b^2}{4}\right]^{\frac{|l-r|}{2}}\left[\frac{1-b}{1+b}\right]^{ \frac{\Jr\, \textrm{sgn}\, (l-r)}{2}} \frac{1+(-1)^{\Jr + |l-r|}}{2}, \label{split_p1}\\
\mathcal{P}_2(l-r|t) &\equiv \rho^{2t} \sum_{l,r=0}^{t} \binom{t}{l} \binom{t}{r} \nu^{l+r}. \label{split_p2}
\end{align}
Using the Gaussian approximation (\ref{Gaussian_approx}), the two split distributions can be approximated as 
\begin{align}
\mathcal{P}_1(\Jr|l-r)  & \stackrel{r-l}{\asymp} \frac{1}{\sqrt{2\pi(1-b^2)|l-r|}}\exp \left[-\frac{(\Jr+(l-r)b)^2}{2(1-b^2)|l-r|} \right] \quad \textrm{for}\ \Jr  + (l-r)b \stackrel{|l-r|}{\sim}  |l-r|^{1/2} ,\\
\mathcal{P}_2(l-r|t) &\asymp \frac{1}{\sqrt{4 \pi t \varDelta^2}} \exp\left[- \frac{(l-r)^2}{4 \varDelta^2 t} \right] \quad \textrm{for}\ l - t \rho,\,  r - t \rho \sim t^{1/2}.
\end{align}
Using the definition of the MGF (\ref{g_def}), converting the sums over $\Jr$ and $l-r$ into integrals, introducing
\begin{equation}
l-r = w \varDelta \sqrt{2t}, \qquad a_\pm \equiv \varDelta t^{1/2} \left(\frac{1-b^2}{2} \lambda^2 \pm b \lambda \right),
\end{equation}
 and evaluating first the integral over $\Jr$ we find
\begin{equation}
G(\lambda|t) \asymp \frac{1}{\sqrt{2\pi}} \sum_{\pm} \int_{0}^{\infty} \dd w \exp \left[ - \frac{w^2}{2} +  \sqrt{2} a_\pm w \right].
\end{equation}
Integrating over $w$ we arrive at the generator of cumulant asymptotics $\mathcal{F}$
\begin{equation}
G(\lambda|t) \asymp \frac{1}{2}\left[e^{a_+^2}\left(1 + \textrm{erf}\, a_+\right) +e^{a_-^2}\left(1 + \textrm{erf}\, a_-\right) \right] \equiv \exp\left[\mathcal{F}^{[b\geq0]}(\lambda)  \right]. \label{Vincent_F}
\end{equation}
From Eq.~(\ref{Vincent_F}) we recover the leading-order cumulant asymptotics (\ref{ref_c2},\ref{ref_c4},\ref{ref_c6},\ref{ref_c8},\ref{ref_c10},\ref{ref_cb0},\ref{ref_cb2},\ref{ref_cb4},\ref{ref_cb6},\ref{ref_cb8},\ref{ref_cb10}) for all $|b|<1$
\begin{equation}
c_n^{[b \geq 0]}(t) \asymp  \frac{\dd^n}{\dd \lambda^n}\Big|_{\lambda = 0} \mathcal{F}^{[b \geq 0]}(\lambda).
\end{equation}

\subsection{Probability distributions}
We now turn our attention to the asymptotics of the exact probability distribution of the integrated current (\ref{P_exact}). To this end we introduce the rescaled current at a scale $\zeta$, $ 0 \leq \zeta \leq 1$
\begin{equation}
\Jcal(t) \equiv t^{-\zeta}\Jr(t).
\end{equation}
where we suppress the dependence of the scaled current on the scaling exponent $\zeta$ to streamline the notation.
Note that to preserve the normalization of the probability distribution of the rescaled current, $\int \mathcal{P}_\zeta(\Jcal|t) \dd \Jcal = 1$, it is necessary to rescale the distribution as
\begin{equation}
\mathcal{P}_\zeta(\Jcal|t) \equiv t^{\zeta} \mathcal{P}(\Jr=\Jcal t^{\zeta}|t) .
\end{equation}
The probability distribution of typical fluctuations on the scale $\zeta = 1/2z$ is independent of time. We accordingly define it as the limit
\begin{equation}
\mathcal{P}_{\textrm{typ}} (j) \equiv \lim_{t \to \infty} \mathcal{P}_{1/2z}(\Jcal=j|t).
\end{equation}

\subsubsection{Normal typical fluctuations for $0<|b|<1$}
For $0<|b|<1$ we have the dynamical exponent $z=1$.
 The cumulants $\kappa_{n}(t)$ of the probability distribution $\mathcal{P}_{1/2}(\Jcal | t)$ are connected to the raw cumulants (\ref{cum_def}) by a rescaling
\begin{equation}
\kappa_n(t) = c_n(t) t^{-n/2}.
\end{equation}
The asymptotics of the cumulants (\ref{cumulant_growth}) for $0<|b| < 1$ implies that all but the second cumulant $\kappa_2(t)$ vanish as $t \to \infty$
\begin{equation}
\lim_{t \to \infty} \kappa_2(t) = 2\varDelta^2b^2, \quad \lim_{t \to \infty} \kappa_{n>2} (t) = 0.
\end{equation}
By the Marcinkiewicz theorem it follows that the probability of the rescaled current on the typical scale is a Gaussian with a variance fixed by the second cumulant
 \begin{equation}
\mathcal P^{[b]}_{\rm typ}(j) = \frac{1}{2\sqrt{\pi} |b| \varDelta} \exp \left[{-\frac{j^2}{4 \varDelta^2 b^2}} \right]. \label{b_pos_dist3}
\end{equation}

\subsubsection{Asymptotic distributions for $b=0$}
We now turn our attention to the unbiased case $b=0$, where the asymptotics (\ref{cumulant_growth}) implies that all even cumulants $\kappa_n(t) = c_n(t) t^{-n/4}$ of the probability distribution $\mathcal{P}_{1/4}(\Jcal|t)$  are non-zero
\begin{equation}
\lim_{t \to \infty} \kappa_{n}(t) = \kappa_n, \quad \kappa_{2n} \neq 0.
\end{equation}
In what follows we consider the scales $ 0 \leq \zeta < 1$.
The dominant contribution to the sum (\ref{P_exact}) comes from two peaks, symmetric with respect to the diagonal $l=r$. 
For $t \to \infty$ the dominant contribution to the sum will not be near the boundary so that we can again apply the Stirling approximation (\ref{binom_approx}) to the binomials in (\ref{P_exact}). Passing from a sum to an integral we must also account for the checkerboard pattern of the right-most sign factor, which we do by introducing an overall factor $\tfrac{1}{2}$. In terms of rescaled variables $x = l/t$, $y=r/t$ the distribution reads
\begin{equation}
t^{-\zeta}\mathcal P^{[0]}_\zeta(\Jcal|t) \asymp \frac{t^{1/2}}{\sqrt{2\pi}^3} \iint_{[0, 1]^2} \frac{e^{t(g-f) + \Jcal t^\zeta h}}{\sqrt{x(1-x)y(1-y)}}  \sqrt{\frac{|x-y|}{(x-y)^2 -\Jcal^2/t^{2(1-\zeta)}}} \Theta \left(|x-y| \geq |\Jcal|/t^{1-\zeta} \right) \dd x \dd y,\\
\end{equation}
where $\Theta(x)$ is the Heaviside function and
\begin{align}
f &= \log \left( \frac{x^x(1-x)^{1-x} y^{y}(1-y)^{1-y}}{\rho^2 \nu^{x + y}}  \right),\\ 
g &= |x-y|\log \left( \frac{|x-y|}{\sqrt{(x-y)^2 - \Jcal^2/t^{2(1-\zeta)}}} \ \right),\\ 
h &= \frac{1}{2}\log \left(\frac{|x-y| - \Jcal/t^{1-\zeta}}{|x-y| + \Jcal/t^{1-\zeta}} \right).
\end{align}
Given the geometry of the integral we again introduce rotated coordinates (\ref{rot_coord})
and anticipating localization of the integral, extend the integration across the entire plane
{\small
\begin{gather}
t^{-\zeta}\mathcal P_\zeta^{[0]}(\Jcal|t) \asymp \frac{t^{1/2}}{\sqrt{2\pi}^3} \iint_{\mathbb{R}^2} \frac{e^{ t(g(s_-)-f(s_+, s_-)) +\Jcal t^\zeta h(s_-)}}{\sqrt{(s_+^2 - s_-^2)(1-\frac{s_+ + s_-}{2})(1 - \frac{s_+-s_-}{2})}}  \sqrt{\frac{|s_-|}{s_-^2 -\Jcal^2/t^{2(1-\zeta)}}} \Theta \left(|s_-| \geq |\Jcal|/t^{1-\zeta} \right) \dd s_+ \dd s_-.
\end{gather}}
Taking into account the Heaviside functions and the two peaks we arrive at
\begin{gather}
t^{-\zeta }\mathcal P^{[0]}_\zeta(\Jcal |t) \asymp \frac{2 t^{1/2}}{\sqrt{2\pi}^3} \int_{\Jcal/t^{1-\zeta}}^\infty \dd s_-   \frac{\sqrt{s_-} e^{tg(s_-) +\Jcal t^\zeta h(s_-)} }{\sqrt{s_-^2 - \Jcal^2/t^{2(1-\zeta)}}} \int_{\mathbb{R}} \dd s_+    \frac{e^{- tf(s_+, s_-)}}{\sqrt{(s_+^2 - s_-^2)(1-\frac{s_+ + s_-}{2})(1 - \frac{s_+-s_-}{2})}}.
\end{gather}
The inner integral can be evaluated by straightforward localization (\ref{Laplace}) around the critical point $s_c^+$
\begin{gather}
\partial_{s_+}\Big|_{s_+ = s_+^c}  f(s_+, s_-) = 0, \quad s_+^c = \frac{2\nu^2 - \sqrt{4\nu^2 + (\nu^2-1)^2s_-^2}}{\nu^2 - 1},
\end{gather}
where the branch of the square root is taken so that $\lim_{\nu \to 1} s_+^c = 1$.  This also ensures that for all $\nu \in \mathbb{R}_+$ we have $ 0<s_+^c <2$.
In the limit of small $s_-$ the critical value of $s_+^c$ becomes
\begin{equation}
s_+^c = \frac{2\nu}{\nu+1} - \frac{\nu^2 - 1}{4\nu} s_-^2 + \mathcal{O} \left( s_-^4 \right) = 2(1-\rho)  - \frac{1-2\rho}{4 \rho(1-\rho)} s_-^2 + \mathcal{O} \left( s_-^4 \right).
\end{equation}
As $t \to \infty$ the maximum in $s_-$  approaches zero so we can expand the functions in the exponent
\begin{align}
g(s_-) \asymp \frac{\Jcal^2}{2s_-} t^{2(\alpha - 1)},  \quad 
f(s_-) \asymp -\frac{\Jcal}{s_-}t^{\alpha-1},
\end{align}
and the integral becomes
\begin{gather}
t^{-\zeta} \mathcal P^{[0]}_\zeta(\Jcal|t) \asymp \frac{1}{ \sqrt{2} \pi \varDelta }  \int_{\Jcal t^{\zeta-1}}^\infty \dd s_-  \, \frac{\sqrt{s_-}}{\sqrt{s_-^2 - \Jcal^2t^{2(\zeta-1)}}} e^{ - t \frac{s_-^2}{4\varDelta^2}  - \frac{\Jcal^2}{2s_-}t^{2\zeta - 1}}. \label{eq10}
\end{gather}
To analyze the localization behavior in the exponential we rescale the integration variable
\begin{equation}
s_- = st^{-\omega}
\end{equation}
so that the expression in the exponential of Eq.~(\ref{eq10}) becomes
\begin{equation}
 \frac{s^2}{4\varDelta^2}t^{1 - 2\omega}  + \frac{\Jcal^2}{2s}t^{2 \zeta  + \omega -1}.
\end{equation}
We equate the two algebraic exponents by setting
\begin{equation}
\omega = \frac{2}{3}(1 - \zeta) \quad \Rightarrow \quad t^{\frac{4\zeta - 1}{3}} \left(\frac{s^2}{4\varDelta^2}  + \frac{\Jcal^2}{2s} \right),
\end{equation}
and, computing the localization, arrive at
\begin{equation}
\mathcal P^{[0]}_\zeta(\Jcal |t) \asymp \frac{t^{\frac{4\zeta - 1}{3}}}{\sqrt{2}\pi\varDelta} \int_{0}^\infty \dd s  \, \frac{1}{\sqrt{s}} \exp \left[{ - t^{\frac{4\zeta - 1}{3}} \left( \frac{s^2}{4\varDelta^2}  + \frac{\Jcal^2}{2s}\right)} \right]. \label{b0_loc_eq}
\end{equation}
We now clearly distinguish between three regimes:
\begin{enumerate}
\item  \emph{Small fluctuations} for $0 \leq \zeta < 1/4$. Here the exponential does not localize, but is instead close to unity on the entire domain, leading to an $\mathcal{O}(1)$ probability.\\
\item  \emph{Typical fluctuations} for $\zeta = 1/4$, where we find a non-gaussian distribution of typical fluctuations
\begin{equation}
\mathcal P^{[0]}_{\textrm{typ}}(j) = \frac{1}{\sqrt{2}\pi\varDelta} \int_{0}^\infty \dd s  \, \frac{1}{\sqrt{s}} \exp \left[ -  \frac{1}{2} \left( \frac{s^2}{2\varDelta^2}  + \frac{j^2}{s}\right) \right].
\end{equation}
A more convenient integral representation is obtained by eliminating the square root factor by a change of variables $s \to u^2$ and extending the integral over the entire real line
\begin{equation}
\mathcal P^{[0]}_{\textrm{typ}}(j) = \frac{1}{\sqrt{2}\pi\varDelta} \int_{-\infty}^\infty \dd u  \, \exp \left[-  \frac{u^4}{4\varDelta^2}  -\frac{j^2}{2u^2} \right]. \label{b0_typical}
\end{equation}
\item \emph{Moderate fluctuations} for $1/4 < \zeta <1$, where the exponential in Eq.~(\ref{b0_loc_eq}) localizes around
\begin{equation}
s_c = \left(|\Jcal|\varDelta\right)^{\frac{2}{3}}
\end{equation}
and after some work gives the rate function in the absence of bias
\begin{equation}
\mathcal P^{[0]}_{\zeta}(j|t)\asymp  t^{\frac{4\zeta - 1}{6}}\frac{\exp\left[-t^{\frac{4 \zeta - 1}{3}} \frac{3}{4} \left( \frac{j^2}{\varDelta}\right)^{2/3}\right] }{\sqrt{3 \pi} (| j| \varDelta)^{\frac{1}{3}}}. \label{rate_function1}
\end{equation}

\end{enumerate}

\subsection{Moderate deviations and the scaled cumulant generating function}
We demonstrate that while the SCGF (\ref{F_res}) does not generate (scaled) cumulants, we can use it to recover the rate function of moderate deviations for $1/2z<\zeta<1$. 
 For $t \to \infty$ we have from the definition of the SCGF (\ref{F_def})
\begin{equation}
e^{t F(\lambda)} \asymp 
 \int \dd \Jcal\, \mathcal{P}_\zeta(\Jcal|t) e^{\lambda\Jcal t^\zeta}. \label{F_P_rel}
\end{equation}
Making an exponential ansatz for the asymptotic probability distribution of the rescaled current
\begin{equation}
\mathcal{P}_\zeta(\Jcal=j|t) \asymp e^{-t^{\nu(\zeta)}I_\zeta(j)},\label{LD_ansatz}
\end{equation}
where $v(\zeta)$ is a `speed' associated to the scale $\zeta$,
we come to
\begin{equation}
e^{t F(\lambda)} \asymp \int \dd \Jcal\, e^{\lambda j t^\zeta -t^{v(\zeta)}I_\zeta(j)}.
\end{equation}
To analyze the localization of this integral we introduce a dynamical rescaling of the counting field
\begin{equation}
\lambda = \eta t^{v(\zeta) - \zeta}, \label{dynamical_rescaling}
\end{equation}
in terms of which the exponential under the integrals is homogeneous in $t$
\begin{equation}
e^{t F(\eta, t)} \asymp \int \dd j\, e^{t^{v(\zeta)}( j \eta  - I_\zeta( j))}. \label{proto_moderate}
\end{equation}
The left hand side of (\ref{proto_moderate}) is to be understood as the limit $t\to \infty$ with $\eta$ held constant.  In this limit, under the assumption $v(\zeta) < \zeta$, the expansions of the SCGF (\ref{Fn0},\ref{F0})  simplify  since higher order terms in $\eta$ are suppressed
\begin{gather}
F^{[b]}(\eta, t) = F^{[b]}(\eta) t^{2 (v(\zeta) - \zeta)}  + \mathcal{O}( t^{3(v(\zeta) - \zeta)}), \qquad F^{[b]}(\eta) \equiv  b^2 \varDelta^2 \eta^2, \label{F_ser0}\\
F^{[0]}(\eta, t) = F^{[0]}(\eta)  t^{4 (v(\zeta) - \zeta)} + \mathcal{O}( t^{6(v(\zeta) - \zeta)}), \qquad F^{[0]}(\eta) \equiv \frac{\varDelta^2 \eta^4}{4}. \label{F_ser1}
\end{gather}

\subsubsection{Moderate deviations}
For $v(\zeta) > 0$ the integral in (\ref{proto_moderate}) can be computed by localization (\ref{Laplace}) and yields
\begin{equation}
e^{t F(\eta, t)} \asymp t^{-v(\zeta)/2} \exp\left( t^{v(\zeta)} \max_{ j} \{\mu  j - I_\zeta( j) \}\right). \label{F_I_iden}
\end{equation}
Inverting (\ref{F_I_iden}) by Lagrange duality, we obtain an expression for the rate function at scale $\zeta$
\begin{equation}
 I_\zeta(j) = \max_{\eta}\{\eta j - F(\eta) \}. \label{rescaled_Legendre}
\end{equation}
By matching the powers of $t$ in the exponents in Eq.~(\ref{F_I_iden}) we find the speeds $v(\zeta)$
\begin{equation}
v^{[b]}(\zeta) = 2 \zeta -1,  \quad
v^{[0]}(\zeta) = \frac{4 \zeta -1}{3}. \label{speeds}
\end{equation}
We observe that the condition $v(\zeta) < \zeta$ is satisfied for $\zeta < 1$, while the `localization condition' $v(\zeta)>0$ is satisfied when $\zeta > 1/2z$, i.e. for
$\zeta > 1/2$ when $|b|>0$  and for  $\zeta > 1/4$ when $b=0$. Provided these two conditions are satisfied Eq.~(\ref{rescaled_Legendre}) gives a ($\zeta$-independent) rate  functions of moderate fluctuations for the range of scales
\begin{equation}
1/4<\zeta^{[0]}<1, \qquad  1/2<\zeta^{[b]}<1.
\end{equation}
Computing the Legendre transforms in (\ref{rescaled_Legendre}) and using (\ref{LD_ansatz}) we exactly recover the moderate deviations for $b=0$ (\ref{rate_function1}). For $0<|b|<1$ we extract the rate function of moderate deviations
\begin{equation}
\mathcal P^{[b]}_\zeta(j|t)\asymp \frac{t^\frac{2\zeta-1}{2}}{2\sqrt{\pi} |b| \varDelta} \exp\left[{-t^{2\zeta - 1}\frac{j^2}{4 \varDelta^2 b^2}}\right] . \label{b_pos_dist2}\\
\end{equation}


\subsubsection{Large deviations}
The same analysis also shows why the scale $\zeta=1$, i.e. large deviations, is distinguished. When $\zeta=1$ both scalings (\ref{speeds}) reduce to $v(1)=1$
implying that higher order terms in $\lambda$ of (\ref{F_ser0},\ref{F_ser1}) are not suppressed and the rate function is the Legendre transform of the full SCGF (\ref{F_res}), which we are however unable to compute in closed form.\\

\noindent Since all species move with unit velocity, the current distribution is supported on the interval $[-t, t]$, implying that the rate function on the scale $\zeta = 1$ is finite only on the interval $\mathcal{I} = [-1, 1]$. From the expansion of the SCGF around $\lambda \to \infty$ (\ref{inf_asymp}) we can extract the behavior of the rate function near the endpoints of the interval for $\delta \equiv 1- |j| $ as
\begin{equation}
I_{1}(j) = \log \frac{2}{(1+|b|)  \varDelta^2} +  \delta \log \delta + \mathcal{O}\left(\delta\right).
\end{equation}

\subsection{Large deviation rate function at the free point}
\noindent For $|b |= 1$ we can explicitly compute the Legendre transform of the SCGF to obtain the rate function
\begin{equation}
I^{[1]}_{1}(j) = \max_{\lambda}\{ \lambda j - F(\lambda) \}.
\end{equation}
The maximum is attained at
\begin{equation}
\lambda_c = \log \left[ \frac{ j(1-2\varDelta^2 ) + \sqrt{j^2(1-4\varDelta^2)+4\varDelta^4}}{2\varDelta^2(1- j)} \right],
\end{equation}
where we find an even rate function
{\small
\begin{equation}
I^{[1]}_{1}(j) =  j \log \left[ \frac{ j(1-2\varDelta^2 ) + \sqrt{[ j(1-2\varDelta^2)]^2+4\varDelta^4(1-  j^2)}}{2\varDelta^2(1- j)} \right] - \log \left[ \frac{1-2\varDelta^2 + \sqrt{[ j(1-2\varDelta^2)]^2+4\varDelta^4(1- j^2)}}{1- j^2} \right].
\end{equation}
}

\section{Numerical estimation of current fluctuations}
Efficient numerical estimations are facilitated by splitting of the exact current distribution as (\ref{p_split})
\begin{equation}
\mathcal{P}(\Jr|t) = \sum_{l-r}\mathcal{P}_1(\Jr|l-r) \mathcal{P}_2(l-r|t),
\end{equation}
where the distributions are given by (\ref{split_p1},\ref{split_p2}).
To independently sample random currents we follow a two step protocol. First we calculate the direction and amount of transfered charge  at time $t$
\begin{equation}
d_t=l_t-r_t,
\end{equation}
by sampling $l,r$ from ($t$-dependent) binomial distributions
\begin{equation}
p(l) = \mathcal B\left(l|t, \rho\right), \qquad p(r) = \mathcal B\left(r|t, \rho\right),
\end{equation}
where $\mathcal B(k|n,\rho) \equiv \binom{n}{k} \rho^k (1-\rho)^{n-k}$.
Next we sample  the number of $+$ particles out of $|d|$ particles $n^{+}_{|d|}$ 
from the binomial distribution
\begin{equation}
p(n^+) = \mathcal B\left(n^+ \big ||d|, (1+b)/2\right),
\end{equation}
leading to the total transfered current
\begin{equation}
\Jr_t=2n^{+}_{|d_t|}\text{sgn}(d_t)-d_t.
\end{equation}
The estimated current distribution $\tilde{\mathcal{P}}$ is obtained by computing histograms
of ${\rm N}_{\rm sample}$ independently drawn currents
\begin{equation}
\tilde{\mathcal{P}}(\Jr|t) = \frac{{\rm N}_\delta (\Jr_t)}{{\rm N}_{\rm sample}},
\end{equation}
where ${\rm N}_\delta (\Jr)$ is the number of currents sampled in a window of width $\delta$ centered on $\Jr$. The estimated moments of the distributions $\tilde G^{(n)}$ are computed as
\begin{equation}
\tilde G^{(n)}(t) = \sum_k (k \delta)^n  \tilde{\mathcal{P}}(k \delta |t).
\end{equation}
Cumulants are estimated from current distribution's moments according to (\ref{Faa_log})
\begin{equation}
\tilde c_n(t) =   \sum_{r=1}^n (-1)^{r-1} (r-1)! B_{n, r}\left(0, \tilde G^{(2)}(t), 0, \tilde G^{(4)}(t), \ldots, \tilde G^{(n-r+1)}(t) \right). \label{c_n_estimate}
\end{equation}
Monte Carlo errors
are estimated from the variance after spliting the total sample into
$10^{3}$ independent subsamples.

\section{Dynamical Lee--Yang theory in equilibrium}
\label{sec:LY}
We give a brief overview of Lee--Yang theory as applied to the MGF of time-integrated current (\ref{J_def}). We start with a general outline of the theory for equilibrium systems.  
\subsection{Lee--Yang zeros}
 Consider the definition of the moment generating function (\ref{g_def})
\begin{equation}
G(\lambda|t)  = \sum_{\Jr} \mathcal{P}(\Jr|t) e^{\lambda \Jr}. \label{LY_Gdef}
\end{equation}
The probability distribution is by definition non-negative, $\mathcal{P}(\Jr|t) \geq 0$, implying that the MGF cannot (for finite times) have zeros on the real axis $\lambda \in \mathbb{R}$. The main idea of the Lee--Yang theory is to instead consider Eq.~(\ref{LY_Gdef}) for complex values of the counting field
\begin{equation}
\lambda \in \mathbb{C},
\end{equation} 
and study the properties of the moment generating function in the complex plane. 
The probability distribution of the current is real implying an involutive symmetry of the MGF in the complex plane
\begin{equation}
G(\lambda|t) = \overline{G(\overline \lambda|t)}, \label{real_sym}
\end{equation}
where $\overline \bullet$ denotes complex conjugation.  In equilibrium the condition of detailed balance, $\mathcal{P}(\Jr|t) = \mathcal{P}(-\Jr|t)$, implies an additional involutive symmetry of the MGF
\begin{equation}
G(\lambda|t) = G(-\lambda|t). \label{inversion_sym}
\end{equation}
 Of particular interest are the (time-dependent) zeros $\lambda_j(t)$ (Lee--Yang zeros) of the MGF, since they are the singularities of the CGF (see Eq.~\ref{CGF_def})
\begin{equation}
G(\lambda_j|t) = 0. \label{LY_def}
\end{equation}
To avoid cluttering the notation we often suppress the time dependence of the zeros, $\lambda_i\equiv\lambda_i(t)$. 
From the symmetries (\ref{real_sym},\ref{inversion_sym}) it follows that the zeros come in (possibly degenerate) quartets 
\begin{equation}
(\lambda_i, \overline{\lambda}_i, -\lambda_i, -\overline \lambda_i). \label{LY_quartet}
\end{equation}
Given that the probability distribution of the current $\mathcal{P}(\Jr|t)$ is supported on the interval $[-t, t]$ the MGF is a rational function in the variable  $e^\lambda$. Consequently the MGF can be factorized as
\begin{equation}
G(\lambda |t) = g(\lambda |t)\prod_{i=1}^{2t} \left(1-\lambda/\lambda_i \right), \quad g(\lambda |t) = e^{-\lambda t} \prod_{i=1}^{2t} \frac{e^\lambda - e^{\lambda_j}}{1 - e^{\lambda_i}} \frac{1}{1 - \lambda/\lambda_i}, \label{G_factorized}
\end{equation}
where the normalization is fixed by the normalization of the MGF (see (\ref{g_norm})). By construction, the function $g(\lambda|t)$ does not have any zeros within the strip $ Z = \{x+ \ii y\ |\  x \in \mathbb{R}, -\pi < y < \pi \}$ and is an analytic function near the origin. We refer to the zero of the MGF closest to the origin in the first quadrant as the \emph{leading} Lee--Yang zero $\lambda_1$
\begin{equation}
|\lambda_1| \leq |\lambda_i| \textrm{ for } 1< i \leq 2t, \quad  0 \leq \textrm{arg}\, \lambda_1 \leq \pi/2. \label{leading_LY}
\end{equation}
Noting that the modulus of the leading Lee--Yang zero is by definition equal to the convergence radius $r$ of the CGF we denote
\begin{equation}
\lambda_1(t) = r(t) e^{\ii\varphi(t)}, \qquad r, \varphi \in \mathbb{R}.
\end{equation}

\subsection{Critical points}
As noted above, the MGF cannot vanish for $\lambda \in \mathbb{R}$ for finite times $t$, implying that the cumulant generating function (\ref{CGF_def})
is a real analytic function. From the factorization (\ref{G_factorized}) it can be expressed as
\begin{equation}
\log G(\lambda |t) = \log g(\lambda|t) + \sum_{i=1}^{2t}  \log \left(1 - \lambda/\lambda_i \right). \label{F_factorized}
\end{equation}
 On the other hand as $t \to \infty$ the Lee--Yang zeros can approach a \emph{critical} point\footnote{There can be more than one critical point. For simplicity we confine the discussion to only one.} $\lambda_c$  on the real axis
\begin{equation}
\lambda_i(t) \xrightarrow{t \to \infty} \lambda_c \in \mathbb{R},
\end{equation}
in the vicinity of which the CGF has a logarithmic singularity
\begin{equation}
\log G(\lambda|t) =   \log (\lambda - \lambda_c) + \mathcal{O}(|\lambda-\lambda_c|^{-1}). \label{dpt_def}
\end{equation}

\subsection{Cumulants, Lee--Yang zeros and critical points}
We now establish a link between the higher cumulants and the dynamics of the leading Lee--Yang zero. We use this to demonstrate an intimate connection between critical points and diverging cumulants.
From the definition (\ref{cum_def}) and the factorization (\ref{G_factorized}) it follows that cumulants are related to Lee--Yang zeros as 
\begin{equation}
c_n(t) = - (n-1)! \sum_{i=1}^{2t} \lambda_i^{-n}(t) + \frac{\dd^n }{\dd \lambda^n} \Big|_{\lambda=0} \log g(\lambda|t). \label{cum_split}
\end{equation}
 We observe that the expression for the cumulants contains a term coming from Lee--Yang zeros and a derivative of a function analytic in the strip $Z$. From asymptotic analysis it follows by Darboux theorem \cite{Dingle}  that for higher-order cumulants the expression (\ref{cum_split}) is dominated by the first term owing to universal oscillations of higher derivatives \cite{Berry}
\begin{equation}
c_{n}(t) \stackrel{n}{\asymp} - (n-1)! \sum_{i=1}^{2t} \lambda_i^{-n}(t).\label{cum_split}
\end{equation}
The expression (\ref{cum_split}) is in turn dominated by the leading Lee--Yang zero (\ref{leading_LY}). Assuming an absence of accidental symmetries, there is a quartet of zeros equidistant from the origin (\ref{LY_quartet}) that determines higher-order cumulants
\begin{equation}
c_{n}(t) \stackrel{n}{\asymp} -2(1 + (-1)^n)(n-1)! \frac{\cos(n \varphi(t))}{|r(t)|^{n}}. \label{LeeYang_cum}
\end{equation}
Note that $c_{2n-1}(t)=0$ already follows trivially from the symmetry (\ref{inversion_sym}). We now use the relation (\ref{LeeYang_cum}) to extract the dynamics of the leading Lee--Yang zero from the knowledge of cumulants. To this end, we define the ratios of (even\footnote{In equilibrium the ratios of subsequent cumulants are trivially either zero or divergent.}) cumulants 
\begin{equation}
R_n^\pm(t) = \frac{c_{n\pm2}(t)}{c_{n}(t)}. \label{r_def}
\end{equation}
Using (\ref{LeeYang_cum}) we obtain the identities (for large $n$)
\begin{align}
\frac{R_n^+}{(n+1)n} r^4 + (n-1)(n-2)R_n^- &= 2 r^2 \cos (2 \varphi),\\
\frac{R_{n+2}^+}{(n+3)(n+2)} r^4 + (n+1)n R_{n+2}^- &= 2 r^2 \cos (2 \varphi),
\end{align}
which can be cast as a matrix system
\begin{equation}
\begin{bmatrix}
1 & - \frac{R_n^+}{(n+1)n}\\
1 & - \frac{R_{n+2}^+}{(n+3)(n+2)}
\end{bmatrix}
\begin{bmatrix}
\lambda_1^2 + \overline \lambda_1^2\\
|\lambda_1|^4
\end{bmatrix} =
\begin{bmatrix}
(n-1)(n-2)R_n^-\\
(n+1)n R_{n+2}^-
\end{bmatrix} \label{matrix_system2}.
\end{equation}
The solution of (\ref{matrix_system2}) for $n \gg 1$ uniquely determines the location of the leading Lee--Yang zero $\lambda_1(t)$ in terms of higher order cumulants $c_n = c_n(t)$
\begin{equation}
r(t) \stackrel{n}{\asymp} n\sqrt[4]{\frac{c_n^2- c_{n+2}c_{n-2} }{c_{n+2}^2- c_{n+4}c_{n}}}, \qquad
\varphi(t) \stackrel{n}{\asymp} \frac{1}{2} \arccos \left(\frac{c_{n+2}c_n- c_{n+4}c_{n-2} }{2\sqrt{(c_{n+2}^2- c_{n+4}c_{n})(c_{n}^2- c_{n+2}c_{n-2})}} \right). \label{LeeYang_sol}
\end{equation}
Now consider a scenario when all cumulants at large times scale as $c_n(t) \sim t^{\gamma n+ \delta}$,  $\gamma,  \delta \in \mathbb{R}, \gamma>0$, with generic amplitudes. From (\ref{LeeYang_sol}) it follows
\begin{equation}
r(t) \sim t^{-\gamma }, \label{LY_asymp}
\end{equation}
and the Lee--Yang zero approaches the origin as $t \to \infty$, implying a critical point (\ref{dpt_def}) with $\lambda_c = 0$.

\begin{figure}[h!]
\centering	
\includegraphics[width=0.8\columnwidth]{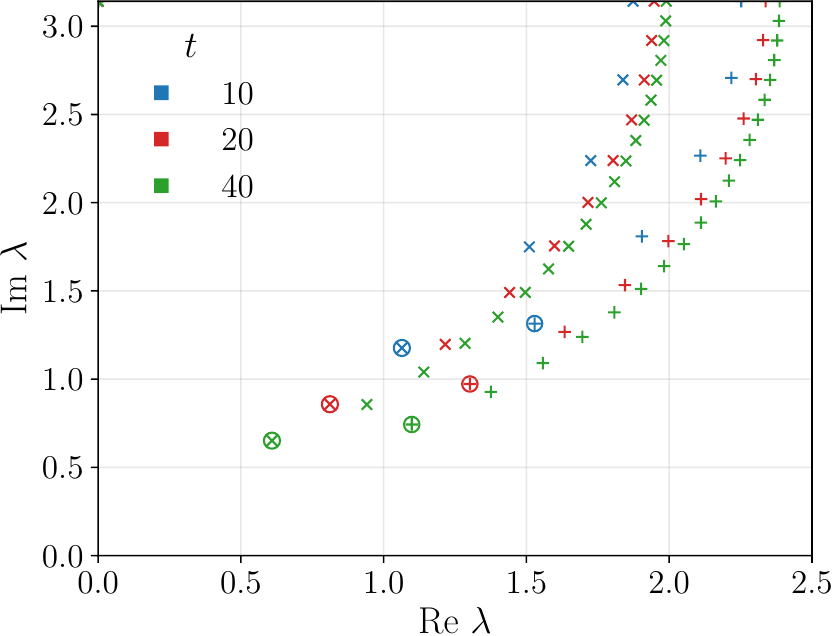}
\caption{Complex Lee-Yang zeros $\lambda_i$  (\ref{LY_def}) of the MGF of the charged hard-core gas (\ref{g_exact}) in the first quadrant $0 \leq \textrm{arg}\ \lambda_i \leq \pi/2$   at three different times $t$ for $b=0$ (+ symbols) or $b=0.5$ ($\times$ symbols) and $\rho=0.5$. The zeros for $b=0.5$ approach the origin faster, in agreement with (\ref{HPG_LY}). Circles show the leading Lee-Yang zeros $\lambda_1$ at different times, for both values of $b$.}
\label{fig_LY}
\end{figure}

\subsection{Leading Lee--Yang zero of the charged hard-core gas}
We are now in a position to apply Lee--Yang analysis to the charged hard-core gas. Taking the cumulant asymptotics (\ref{cumulant_growth}) for $n>2$ and using (\ref{LY_asymp})
we immediately obtain asymptotics of the modulus of the leading Lee--Yang zero 
\begin{equation}
r^{[b]}(t) \sim t^{- 1/2}, \quad r^{[0]}(t) \sim t^{- 1/4}. \label{HPG_LY}
\end{equation}
The result (\ref{HPG_LY}) indicates a critical point at the origin, $\lambda_c=0$, in the charged hard point gas irrespectively of the bias $|b|<1$. For small times, the Lee-Yang zeros are shown in Figure~\ref{fig_LY}.

We emphasize that while the convergence radius of the CGF vanishes around the origin as $t \to \infty$ (for all $|b|<1$) this is not inconsistent with a finite radius of convergence of the SCGF  (\ref{F_res}) for $b=0$ owing to the non-commutativity of limits (\ref{non_commute}).

\section{First order phase transition}
\begin{figure}[h!]
\centering	
\includegraphics[width=\columnwidth]{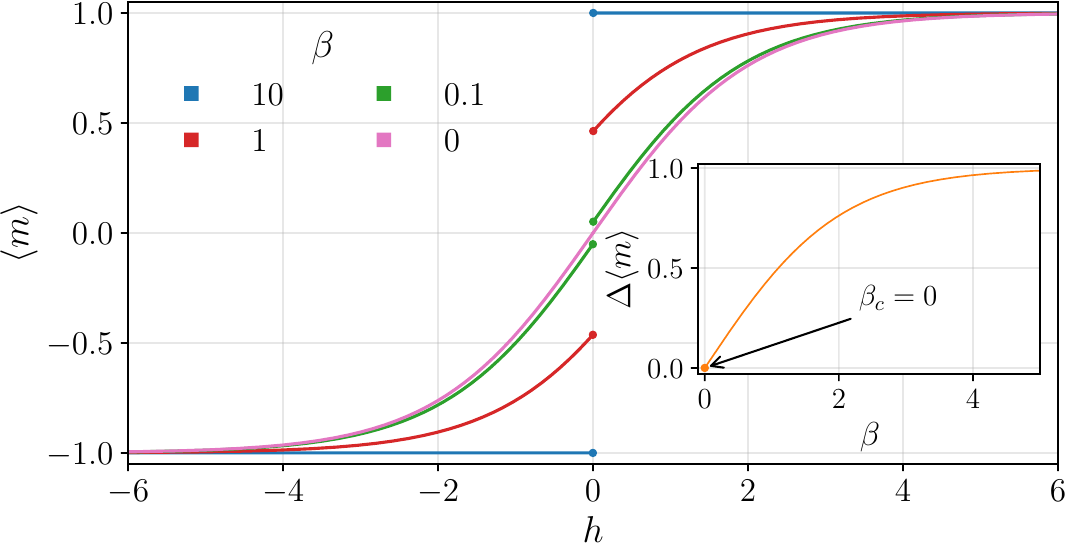}
\caption{Average magnetization $\langle m \rangle$ (\ref{avg_m}) as a function of $h$ at four values of $\beta$. Circles indicate the left/right limits $\lim_{h \to 0^\pm} \langle m \rangle$. (inset) Jump in the average magnetization $\Delta \langle m \rangle \equiv \tfrac{1}{2} \left( \mathcal M|_{h=0^+} - \mathcal M |_{h=0^-} \right)$ as a function of $\beta$. Circle indicates the critical point $\beta_c = 0$. Parameters:  $\nu = 1$}
\label{fig_transition}
\end{figure}

In this section we show that the model exhibits a line of first order phase transitions, by considering the bias $b$ as a control parameter.\\
Reparametrizing the pair of parameters $(\lambda, b)$ in terms of 
\begin{align}
\beta(\lambda, b) &\equiv \log \sqrt{\mu_+ \mu_-} = \log \sqrt{\cosh^2 \lambda - b^2 \sinh^2 \lambda},\label{beta_param}\\
h(\lambda, b) &\equiv \log \sqrt{\mu_+/ \mu_-} = \log \sqrt{\frac{\cosh \lambda +  b \sinh \lambda}{\cosh \lambda -  b \sinh \lambda}}, \label{h_param}
\end{align}
while also introducing a magnetization $M = l-r$, the expression for the MGF (\ref{G_binom}) can be expressed as
\begin{equation}
G(\beta, h|t) = \sum_{M=-t}^{t} e^{\beta|M| + h M}   P(M|t) , \label{G_thermo}
\end{equation}
where  $P(M|t)$ is the probability distribution of $M$ at time $t$
\begin{equation}
P(M|t) = \rho^{2t}\sum_{\substack{l, r = 0,\\ l-r = M}}^{t} \binom{t}{l} \binom{t}{r} \nu^{l+r}. \label{Pm}
\end{equation}
The normalization (\ref{g_norm}), transcribed into the language of $(\beta, h)$, ensures that $P(M|t)$ is normalized to unity
\begin{equation}
\sum_{M=-t}^{t} P(M|t) = 1.
\end{equation}
The expressions (\ref{G_thermo},\ref{Pm}) can be formally identified with a thermodynamic partition sum of a Curie--Weiss type (i.e. mean-field) model of magnetism (see also \cite{Flindt_PRR}), where $t$ plays the role of system size. Introducing rescaled magnetization as $m = M/t$ and sending $t \to \infty$ we can reuse the result (\ref{g_res}) to extract
\begin{equation}
G(\beta, h|t) \asymp e^{t f(x_\gamma, y_\gamma)},
\end{equation}
which gives the average magnetization and the average of it absolute values by taking logarithmic derivatives with respect to $h$ or $\beta$ respectively
\begin{align}
\langle m \rangle &\equiv t^{-1}\partial_h \log G(\beta, h|t),\\
\langle |m| \rangle &\equiv t^{-1}\partial_\beta \log G(\beta, h|t). 
\end{align}
Using the explicit form of $f$ (\ref{saddle_f}) and (\ref{max1},\ref{max_func}), while noting that $\gamma = {\rm  sgn}\, (\lambda b) = {\rm  sgn}\, h$ , we find for $t \to \infty$\footnote{Note that the limit $t \to \infty$ commutes with taking partial derivatives for $\beta \neq 0$, see Section~\ref{sec:LY}.}
\begin{align}
\langle m \rangle &= {\rm sgn}(h)\, \mathcal{M}, \label{avg_m}\\
\langle |m| \rangle &=  \mathcal{M}. 
\end{align}
where
\begin{equation}
\mathcal{M}(\beta, h) = \frac{\sinh(\beta + |h|)}{\cosh(\beta + |h|)  + (\nu+\nu^{-1})/2}.
\end{equation}
Observing that  $\mathcal{M}(\beta, 0) > 0$  for $\beta > 0$, while $\mathcal{M}(0, 0) =0$ it follows from (\ref{avg_m}) that the average magnetization jumps as $h$ is varied across the origin for $\beta > 0$, while it varies continuously for $\beta = 0$ (see Figure~\ref{fig_transition}). This indicates a line of first order phase transitions (in the parameter $h$) which terminates at a critical point $\beta_c = 0$.

We note that the preceding discussion is exactly the behavior observed in localizing the integral (\ref{g_int}) (cf. discussion below Eq.~(\ref{max_func})), translated into a `thermodynamic' language using the parametrization (\ref{beta_param},\ref{h_param}). Moreover we observe that the appearance of the critical point at $\beta_c = 0$ is consistent with the Lee-Yang analysis in Section \ref{sec:LY}.

\end{document}